\documentclass[sigconf,10pt,nonacm]{acmart}
\usepackage{amsmath}
\usepackage{graphicx}
\usepackage{xspace}
\usepackage{indentfirst}
\usepackage{algorithm}
\usepackage{algpseudocode}
\usepackage{multicol}
\usepackage{caption}
\usepackage{makecell}
\usepackage{subcaption}
\usepackage{xspace}
\usepackage{enumitem}
\usepackage{mdframed}
\usepackage{mathtools}

\DeclareFontFamily{U}{stix2bb}{}
\DeclareFontShape{U}{stix2bb}{m}{n} {<-> stix2-mathbb}{}

\newcommand{\sysname}{Opus\xspace}

\providecommand{\ie}{\emph{i.e.,} }
\providecommand{\eg}{\emph{e.g.,} }

\providecommand{\etc}{\emph{etc.}}      
\providecommand{\myparab}[1]{\noindent\textbf{#1} }

\providecommand{\myparab}[1]{\smallskip\noindent\textbf{#1} }
\providecommand{\alltoall}{\textsc{AllToAll}\xspace}
\providecommand{\allreduce}{\textsc{AllReduce}\xspace}
\providecommand{\allgather}{\textsc{AllGather}\xspace}

\providecommand{\sendrecv}{\textsc{Send/Recv}\xspace}
\providecommand{\reducescatter}{\textsc{ReduceScatter}\xspace}
\newlist{compactitem}{itemize}{1}
\setlist[compactitem,1]{label=\textbullet, left=0pt, itemsep=1pt, topsep=1pt, parsep=0pt, partopsep=0pt}

\usepackage{titlesec}[small,compact] 
\titlespacing*{\section}{0pt}{0.1\baselineskip}{0.2\baselineskip}
\titlespacing*{\subsection}{0pt}{0.1\baselineskip}{0.2\baselineskip}
\title{Photonic Rails in ML Datacenters}
\author{Eric Ding}
\affiliation{%
  \institution{Cornell University}%
  \city{Ithaca, NY}
  \country{USA}%
}
\author{Chuhan Ouyang}
\affiliation{%
  \institution{Cornell University}%
  \city{Ithaca, NY}
  \country{USA}%
}
\author{Rachee Singh}
\affiliation{%
  \institution{Cornell University}%
  \city{Ithaca, NY}
  \country{USA}%
}

\begin{abstract}
Rail-optimized network fabrics have become the de facto datacenter scale-out fabric for large-scale ML training. However, the use of high-radix electrical switches to provide all-to-all connectivity in rails imposes massive power, cost, and complexity overheads. We propose a rethinking of the rail abstraction by retaining its communication semantics, but realizing it using optical circuit switches. The key challenge is that optical switches support only one-to-one connectivity at a time, limiting the fan-out of traffic in ML workloads using hybrid parallelisms. We introduce \emph{parallelism-driven rail reconfiguration} as a solution that leverages the sequential ordering between traffic from different parallelisms. We design a control plane, \sysname, to enable time-multiplexed emulation of electrical rail switches using optical switches. More broadly, our work discusses a new research agenda: datacenter fabrics that co-evolve with the model parallelism dimensions \emph{within} each job, as opposed to the prevailing mindset of reconfiguring networks before a job begins.\end{abstract}

\begin{document}

\maketitle

\vspace{-0.5cm}
\section{Introduction}
\begin{figure}[h!]
    \centering
    \includegraphics[width=1\linewidth]{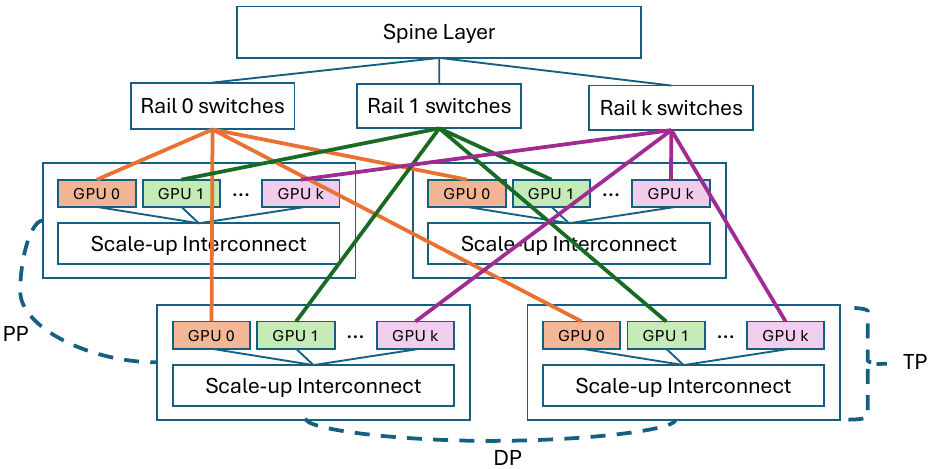}
    \caption{Rail-optimized fabrics. We propose to replace packet switches (shown as Rail 0, Rail 1 \etc) with optical circuit switches. We make the case for retaining the \emph{illusion of full connectivity} between GPU ranks connected to the same optical rail switch using in-job reconfiguration.}
    \label{fig:rail}
\end{figure}

The design of datacenter interconnect fabrics has a significant impact on the scalability and efficiency of large-scale machine learning (ML) systems. A wide range of general datacenter fabric designs have been proposed in the past decade, including electrical~\cite{vl2, jupiter-rising, harsh2020spineless,agarwal2024harmony, singla2014high} and photonic networks~\cite{farrington2010helios,cthrough,firefly,projector,chen2017enabling}. More recently, the rapid growth of ML training workloads has driven a shift toward ML-centric datacenter fabric designs~\cite{jouppi2023tpu, tpuresilience, sipac, wang2023topoopt, lumorph, lumion, sipml}. Among the many proposals, one datacenter topology has seen broad adoption: the rail-optimized fabric~\cite{nvidia-rail-optimize,wang2024rail, gherghescu2024ve}. The rail fabric explicitly aligns with the communication patterns of hybrid parallelisms in ML workloads by wiring together ``rails'' of GPUs---sets of GPUs with identical ranks across multiple high-bandwidth (or \emph{scale-up}) domains, forming the \emph{scale-out} domain (Fig.~\ref{fig:rail}). This design can achieve congestion-free communication for common collectives like \allreduce and \allgather in distributed ML pipelines~\cite{nvidia-rail-optimize}.

But this performance comes at a steep cost. To ensure contention-free communication across rails, rail fabrics are significantly over-provisioned in electrical switch bandwidth \cite{wang2024rail}. Each rail switch connects GPUs of the same rank in all scale-up domains, resulting in networks built from high-radix packet switches. These switches are not only costly, but also power-hungry. Switch ASIC processing, transceiver electrical-optical conversions, and multi-tier Clos, all contribute to the energy and complexity burden of the fabric~\cite{broadcom_scip_2021, nvidia_spectrum-x_2025, jupiter-evolving}. Ironically, the electrical rail design achieves congestion freedom by brute-force overcapacity, even though ML communication patterns are structured and predictable.

This paper asks whether it is possible to retain the desirable properties of rail-optimized fabrics while dramatically improving their energy and cost efficiency. Rather than redesigning the datacenter topology~\cite{wang2023topoopt, lumorph, lumion, liao2025mfabric, slimfly,jellyfish}, we pursue a different direction to answer the question: we propose to replace electrical packet switches in the rail with reconfigurable optical circuit switches (OCSes) which consume a magnitude lower power than their electrical counterparts~\cite{sipac}. 
We call the resulting design a \emph{photonic rail-optimized fabric}. Our proposal draws inspiration from recent successes in optical ML fabrics~\cite{jouppi2023tpu, tpuresilience} but departs from them by preserving the already-successful rail design, while fundamentally changing how data is switched within it.

However, the shift from packet switching to circuit switching in rails is not straightforward. Electrical rail switches enable \emph{all-to-all} packet-level connectivity among GPUs in the same rail, whereas OCSes only offer one-to-one circuit connectivity (one GPU to another GPU) at a given time. This limits the node degree of each GPU rank and breaks a key invariant of rail-optimized designs--full connectivity among the same-rank GPUs across all scale-ups (Fig.~\ref{fig:rail}).
Without full connectivity, hybrid ML parallelisms will become inefficient or even infeasible on rail fabrics. The core technical challenge, then, is: \emph{how to retain the abstraction of seamless rail communication despite the limitations of photonic switching?}

One key observation is that most communication in distributed ML is predictable and structured \cite{gangidi2024rdma, wang2023topoopt}. Collectives are issued in known sequences, organized by parallelism type (\eg tensor, data, pipeline)~\cite{nvidia_nccl}. We exploit this structure to break the illusion of requiring full connectivity by \emph{reconfiguring the photonic fabric between collectives within the job}. To our knowledge, this is the first proposal to reconfigure fabrics between ML collectives, while other proposals reconfigure once prior to the job start~\cite{wang2023topoopt, tpuresilience, jouppi2023tpu}, or reconfigure during workloads specifically for one type of parallelism~\cite{liao2025mfabric}.


We consider the feasibility of in-job reconfiguration that adapts the fabric to each parallelism phase, without disrupting the ML stack or requiring bespoke OCSes. To realize the vision of photonic rails, we propose a novel control layer in ML software stacks, \sysname. The collective communication libraries will act as clients to \sysname and issue provisional intents to communicate. \sysname will interface with the traditional network controller to orchestrate rail reconfiguration in response to such intents. Our early experiments show that \sysname can reduce networking infrastructure cost by $70\%$ and power consumption by $96\%$ while only incurring $3\%$ increase in iteration time over a network with electrical rails.


\section{Communication across Parallelisms}

\begin{table}[H]
    \centering
    \footnotesize
    \begin{tabular}{ccc}
        \hline
        \textbf{Model size} & \textbf{Compute ($N$ GPUs)} & \textbf{Practices}\\
        \hline
        Small ($< 10$B) & $N \leq 8$ & TP or DP \\
        Large ($> 10$B) & $8 < N \leq 512$ & \footnotesize{TP \& PP, TP \& DP, or DP} \\
        Large ($> 10$B) & $512 < N \leq 1024$ & DP \& PP, or DP \& TP \\
        Large ($> 10$B) & $N > 1024$ & TP, DP \& PP \\
        \hline
    \end{tabular}
    \caption{Rule-of-thumb LLM parallelism strategies \cite{ultrascale_playbook}.}
    \label{tab:parallel_strategy}
\end{table}

\begin{table*}[h!]
    \centering
    \footnotesize
    \begin{tabular}{cccc}
    \hline
    \textbf{Parallelism}    & \textbf{Memory reduction} & \textbf{Compute reduction} & \textbf{Communication type and frequency}\\
    \hline
    \hline
    DP        & \makecell{gbs/dp} & \makecell{gbs/dp} & \makecell{bwd AR per layer/per model}  \\
    \hline
    FSDP                     & \makecell{gbs/dp, params/dp} & \makecell{gbs/dp} & \makecell{fwd AG, bwd RS per layer/model} \\
    \hline
    TP      & \makecell{params/tp, grads/tp, optims/tp} & \makecell{params/tp} & \makecell{fwd bwd AR per operator} \\
    \hline
    TP \& SP     & \makecell{params/tp, grads/tp, optims/tp, activs/tp} & \makecell{params/tp, activs/tp} & \makecell{fwd bwd AG\&RS per operator} \\
    \hline
    CP     & \makecell{kv\_cache/cp, seq/cp} &  \makecell{seq/cp} & \makecell{fwd AG bwd RS per layer} \\
    \hline
    PP   & \makecell{params/pp, grads/pp, optims/pp, activs/pp} &  \makecell{params/pp}  & \makecell{fwd bwd \sendrecv per microbatch} \\
    \hline
    EP      & \makecell{experts/ep} &  \makecell{experts/ep} & \makecell{fwd bwd \alltoall per layer} \\
    \hline
    \end{tabular}
    \caption{\small{Characteristics of different parallelism strategies \cite{liang2024torchtitan}. 
        gbs: global batch size.
        dp: data parallel degree.
        seq: sequence length.
        fwd: forward pass.
        bwd: backward pass.
        AR: \allreduce.
        AG: \allgather.
        RS: \reducescatter.
        params: model parameter size.
        grads: gradients size.
        optims: optimizer states.
        activs: activation states.
    }}
    \label{tab:paralllelism}
\end{table*}

\begin{figure}[h!]
    \centering
    \includegraphics[width=1\linewidth]{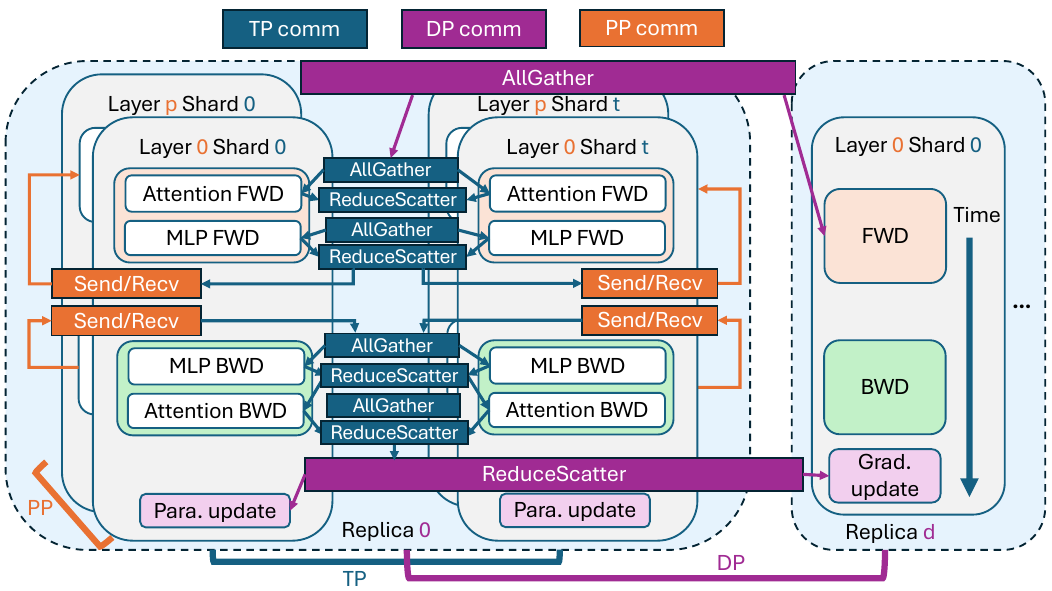}
    \caption{Traffic in a training iteration with 3D parallelism.}
    \label{fig:traffic}
\end{figure}

Training ML models at scale requires a combination of parallelism strategies across thousands of GPUs (Tbl.~\ref{tab:parallel_strategy}) \cite{chu2025scaling, rasley2020deepspeed, liu2024deepseek,shoeybi2019megatron}. To make this feasible, ML systems leverage multiple, co-existing dimensions of parallelism. These include data parallelism (DP, and variants like fully sharded data parallelism or FSDP), pipeline parallelism (PP), tensor parallelism (TP, often combined with sequence parallelism or SP), context parallelism (CP), and expert parallelism (EP) \cite{xing2015petuum, zhao2023pytorch, rasley2020deepspeed, shoeybi2019megatron, korthikanti2023reducing, liu2023ring, fedus2022switch, liu2024deepseek, gherghescu2024ve}. As shown in Tbl.~\ref{tab:paralllelism}, each parallelism axis incurs communication that differs in: (1) data volume--ranging from full model weights in DP to per-layer activations in TP; (2) start time--some collectives occur during the forward pass, others only during backpropagation; (3) frequency--some fire once per layer, others once per micro-batch; and (4) communication pattern--from ring-based \allreduce to high-fanout \alltoall. Importantly, the communication operations from different parallelisms are not ordered arbitrarily: they follow strict dependencies defined by the model’s execution graph (DAG). Fig.~\ref{fig:traffic} illustrates these dependencies in a 3D-parallel training step.



\subsection{Our Proposal: Electrical $\rightarrow$ Optical Rails}
Rail-optimized topologies are gaining traction for scaling ML workloads~\cite{nvidia-rail-optimize}. These topologies organize the network into multiple independent “rails,” where each rail connects GPUs of the same local rank across different scale-up domains, \eg Nvidia DGX~\cite{nvidia_dgx_superpod_2025} or HGX~\cite{nvidia_hgx_platform_2025} nodes. The number of rails equals the number of GPUs in each scale-up domain.

Fig.~\ref{fig:rail} illustrates how a 3D parallelism strategy maps naturally onto a rail-optimized network. Frequent and latency-sensitive TP collectives are confined within the high-bandwidth scale-ups, whereas PP and DP collectives often traverse the slower scale-out network because they occur less frequently and can be overlapped with compute~\cite{qi2023zero,gherghescu2024ve}. The rail abstraction allows these scale-out collectives to occur without oversubscription: every rail provides a dedicated, congestion-free path across domains for a specific GPU rank.

\myparab{Limitations of packet-switched rails.} Despite their performance benefits, today’s rail fabrics suffer from scalability challenges. Optical fibers connect scale-up domains to the network, terminating at transceivers that interface with server NICs and packet switches. Each packet switch introduces optical-electrical-optical (OEO) conversions, adding energy and latency overhead to the data path. These conversions, coupled with the switch ASIC's work---packet queueing, header parsing, and TCAM lookups---consume significant energy \cite{Yeluri2023_power_consumption}. Moreover, while link speeds (\eg 400 Gbps) continue to scale, ASIC processing speed has not kept pace. As a result, packet-switched fabrics now represent a bottleneck in both power efficiency and bandwidth scalability, especially in the face of LLM training jobs that require hundreds of rails and thousands of endpoints \cite{broadcom_cpo}.

\myparab{Replacing rail packet switches with OCSes.} We propose a re-imagined rail-optimized fabric in which each rail is implemented not with electrical packet switches, but with OCSes. These photonic switches can form end-to-end optical paths without OEO conversions, eliminating switch ASICs entirely from the datapath. The result is a dramatic reduction in energy consumption, near-zero datapath latency, and the ability to scale bandwidth without incurring ASIC bottlenecks~\cite{Lumentum2025OCS}. Importantly, our proposal retains the logical structure of the existing rail-optimized topology: the scale-up domains, cabling, and GPU-to-rail mapping, all remain unchanged. There is no multi-tier electrical rail or spine. Instead, each rail becomes a flat, photonic point-to-point fabric. Cross-rank communication can still be supported via forwarding through the high-bandwidth interconnect in scale-up (\eg PXN~\cite{nvidia-pxn}), as explored in prior work~\cite{wang2024rail}. The control plane remains electrical and host-driven.

Mature technologies like MEMS-based OCSes already have characteristics required by this design: millisecond reconfiguration and switch radix in the hundreds~\cite{Lumentum2025OCS, polatis_series7000, liang2024negotiator, liao2025mfabric, ballani2020sirius}. 
We note that our proposal differs from recent silicon photonic architectures proposed by Nvidia and Broadcom~\cite{nvidia_silicon_photonics, broadcom_bcm78909}, which still rely on electrical switching ASICs but use co-packaged optics (CPO) instead of pluggable optical transceivers.


\section{Challenges in Realizing Photonic Rails}
Replacing electrical rail switches with OCSes is non-trivial.
In ML jobs with hybrid parallelisms, each GPU is a member of multiple communication groups --- logical constructs managed by collective communication libraries like NCCL~\cite{nvidia_nccl_communicators}. A single GPU belongs to several groups, each associated with a different parallelism axis. This leads to a high communication degree per GPU. 
For example, the degree requirement is 6 in a 3D--parallel job using ring-based \allreduce, where one GPU has two neighbors per ring. 
Packet-switched rails support this naturally due to their full connectivity. 
However, the number of simultaneous optical circuits per GPU is bounded by its physical degree, \ie the number of network ports. This degree limitation leads to three constraints:

\myparab{C1: Collective algorithm.} Low degree restricts collectives to ring algorithms that are bandwidth-efficient but incur high latency~\cite{taccl}. Latency-optimized strategies like tree-based or recursive-doubling collectives cannot be used~\cite{thakur2003improving, sanders2009two}.

\myparab{C2: Parallelism dimensionality.} The number of parallelisms are constrained as some parallelisms (\eg CP) need to be implemented independently~\cite{liang2024torchtitan}.

\myparab{C3: Bandwidth fragmentation.} Statically partitioning  NIC ports across communication groups allocates only a fraction of NIC bandwidth to each collective.

\myparab{Examples.}
Consider training a model with 3D parallelism (TP, DP, and PP) on DGX H200 nodes, where TP is within the scale-up domain.
Each GPU is mapped to one ConnectX-7 NIC with three port configuration options, \eg one logical 400Gbps port, two logical 200Gbps ports, and four 100Gbps ports \cite{nvidia2025dgx_h200, nvidia_connectx7_datasheet}.
The DP and PP collectives must share the scale-out optical rail network. If 4-port configuration is used by the NIC, two ports must be allocated to each parallelism dimension (for two neighbors in a ring), effectively halving the available bandwidth per group (C3). The low node degree forces ring-based collectives (C1), and adding CP would be infeasible without additional NICs or switching hardware (C2). One workaround is to multiplex parallelisms over shared physical links, but this introduces a new set of problems: forwarding traffic via intermediate GPUs inflates latency and incurs a bandwidth tax~\cite{mellette2017rotornet}. Prior OCS-based ML fabrics have sidestepped these issues by adding multiple NICs per GPU~\cite{wang2023topoopt, sipac} or by placing one parallelism (\eg EP) in the OCS fabric while relying on packet-switched network for other traffic~\cite{liao2025mfabric} --- costly solutions that increase network complexity and fail to support higher-dimensional parallelisms cleanly. Google's TPU cluster~\cite{jouppi2023tpu} uses OCSes to construct a 3D torus before the job starts, mapping parallelisms to x,y,z dimensions but suffers from  C1 and C3~\cite{lumorph, lumion}.


\myparab{Key Insight.}
This naturally raises a provocative question: can we reconfigure the OCSes \emph{during a job} to enable 5D parallelisms? Prior systems have explored microsecond- or nanosecond-scale reconfiguration to enable traffic-aware or traffic-oblivious scheduling in general datacenter networks~\cite{cthrough,projector,farrington2010helios,amir2024shale,shoal, mellette2017rotornet, liang2024negotiator, ballani2020sirius}. However, they are poorly suited to the repetitive and high-volume collective communication patterns of ML workloads~\cite{wang2023topoopt}. We propose to embrace a different unit of adaptation: the collectives themselves. ML collectives occur in well-defined patterns dictated by the model’s computational graph. These patterns naturally create brief \emph{windows} of communication inactivity between phases of parallelism --- ideal time to reshape the rail network topology to match the upcoming collective’s demands.



\subsection{Time Window between Parallelisms}
\label{sec:trace}

To study the window sizes, we run an LLM training workload using TorchTitan~\cite{liang2024torchtitan} on the Perlmutter supercomputer~\cite{nersc_perlmutter_architecture}.
We use 4 nodes connected by Slingshot 11 interconnect fabric~\cite{hpe_cray_ex}. Each node has 4 A100 GPUs inter-connected via NVlink 3.0.
We train Llama3-8B with TP=4 (intra-node), FSDP=2, and PP=2~\cite{grattafiori2024llama}. The PP schedule is 1-forward-1-backward \cite{shoeybi2019megatron}, and the micro-batch size is 2.


\begin{figure}[h]
    \centering
    \includegraphics[width=1\linewidth]{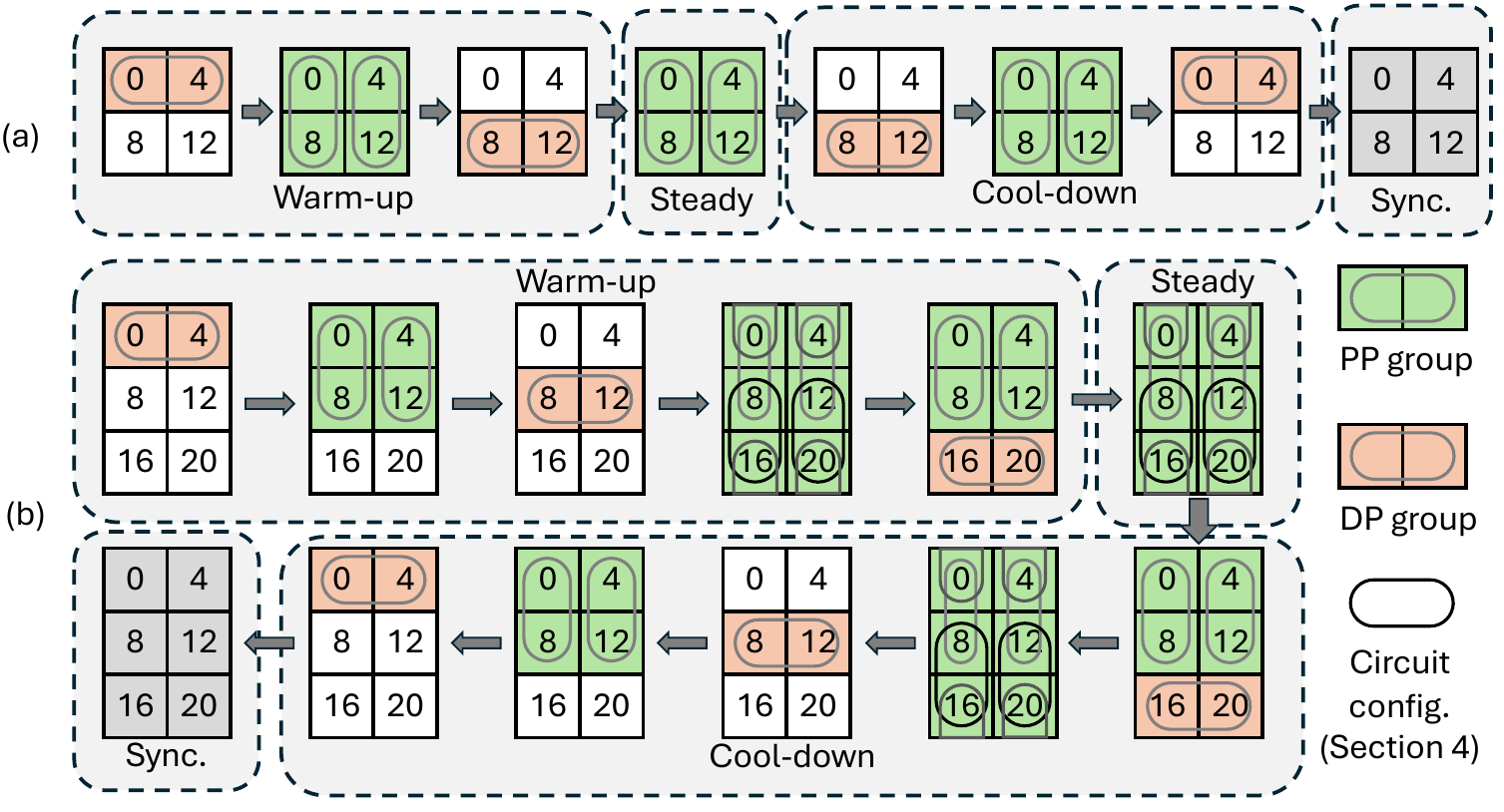}
    \caption{\small{Communication pattern for PP and FSDP in one iteration, split based on the warm-up, steady, and cool-down stages of the pipeline (4 rails in total, only showing rail 0, TP is hidden). (a) PP=2, FSDP=2. (b) PP=3, FSDP=2. }}
    \label{fig:config}
\end{figure}

Fig. \ref{fig:config}(a) shows the communication pattern in rail 0 (same pattern for other rails). Rank 0 first performs stage 0 micro-batch 0 forward pass (overlapped with per-layer \allgather calls to collect the next layer's parameters), and sends the activation to stage 1 hosted by rank 8 through a \sendrecv call along the pipeline dimension. Once the \sendrecv call is finished, rank 8 computes the forward pass, while doing \allgather. Then, it performs backpropagation for micro-batch 0, followed by pipeline \sendrecv. \reducescatter calls are issued after partial gradients are updated. During the optimizer step, several short \allreduce calls are issued for synchronization and numerical robustness~\cite{pascanu2013difficulty, micikevicius2017mixed}. We observe that the DP traffic do not overlap with TP traffic. Fig. \ref{fig:config}(b) shows the pattern for PP=3.
The data dependency between operations, and PyTorch's lazy DTensor operation (\eg the first \allgather call for stage 1 only starts when it receives the activation from stage 0), dictate the sequential order between PP and DP traffic, though collectives from two dimensions are issued in different CUDA streams.



We find that the windows (the arrows in Fig. \ref{fig:config}) are on the order of milliseconds. We define the window as the idle time between two consecutive parallelism phases $P1$ and $P2$, which are two distinctive sets of communication groups: 
$$T_{window} = \min_{comm_j \in P2} T_{comm_j\_start}  - \max_{comm_i \in P1} T_{comm_i\_end},$$ where $comm_j \neq comm_i$ for all $comm_i \in P1$. In addition, 
$$T_{comm_j\_start} = \max_{rank_x \in comm_j} T_{rank_x\_comm_j\_start},$$
where $rank_x$ participates in $comm_j$, since the collective starts only when the slowest rank joins. $T_{comm_i\_end}$ is the end time of the $comm_i$, the same for all participating ranks.

\begin{figure}[ht]
    \centering
    \begin{subfigure}[b]{0.49\linewidth}
        \centering
        \includegraphics[width=\linewidth]{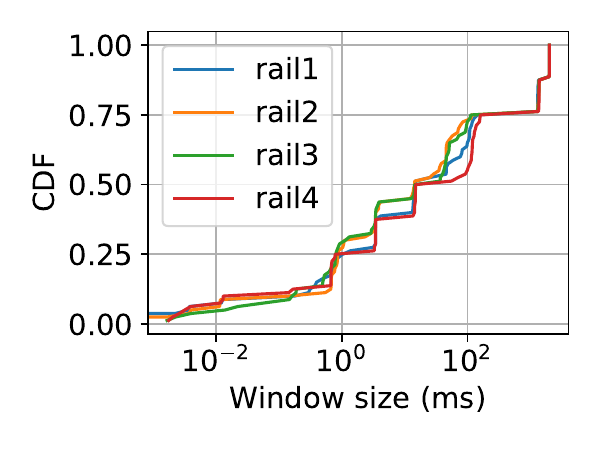}
    \end{subfigure}
    \hfill
    \begin{subfigure}[b]{0.49\linewidth}
        \centering
        \includegraphics[width=\linewidth]{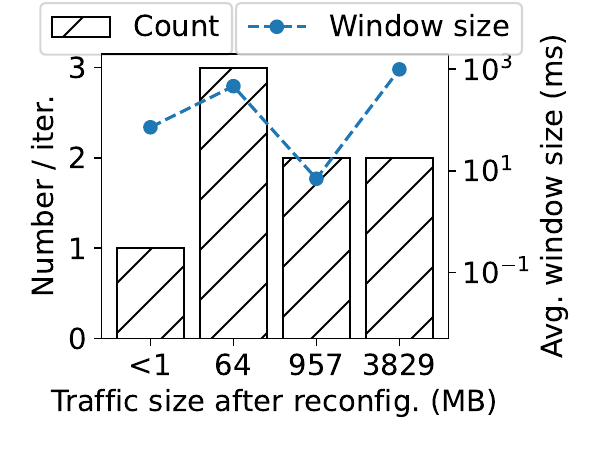}
    \end{subfigure}
    \caption{(a) CDF of window size from 10 iterations. (b) Rail 0 window break-down based on traffic volume after the window and before the next window, in one iteration. <1MB: \allreduce synchronization calls, 64MB: PP \sendrecv, 957MB: DP \allgather, 3829MB: DP \reducescatter).} 
    \label{fig:window}
\end{figure}

Based on the definition, we plot the CDF of the window sizes and categorize the windows according to the total traffic volume in $comm_j$ (communication after the window) in Fig. \ref{fig:window} for the Llama3-8B workload. Our observation is that more than 75\% of the windows are over 1ms long and are similar in size across rails. And the biggest traffic volume (\reducescatter) is preceded by the largest window (1000ms in average). For general workloads, the number of windows in one iteration can be determined by Eq. \ref{eq:wndcnt} (assuming FSDP is used, and TP domain does not exceed scale-up). Using the training configurations reported by~\cite{chu2025scaling}, there are 127 windows over one Llama3.1-405B training iteration, approximately 20 seconds with 1k H100s ($\approx 6$ windows/second)~\cite{nvidia2025llama31_405b_dgxc}.


Our findings indicate that topology reconfiguration can have a minimal impact on the application performance if the reconfiguration delay is on the order of milliseconds, allowing it to be hidden in the windows between parallelisms.

\begin{figure*}[h!]
\footnotesize
\centering
\begin{equation}
\text{Window count} \leq \underbrace{4 \cdot (PP - 1)}_{\parbox{2cm}{\centering PP and FSDP \\ fwd/bwd interleave }} 
+ \underbrace{2 \cdot \frac{n_{\text{layer}}}{PP} - 1}_{\parbox{2cm}{\centering CP/EP and FSDP \\1st microbatch fwd interleave}}
+ \underbrace{4 \cdot n_{\text{microbatch}}}_{\parbox{2cm}{\centering CP/EP and PP \\fwd/bwd interleave}}
+ \underbrace{2 \cdot n_{\text{microbatch}} \cdot ( 2 \cdot \frac{n_{\text{layer}}}{PP} - 1)}_{\parbox{2.5cm}{\centering CP and EP \\fwd/bwd interleave}}
+ \underbrace{4}_{\parbox{2.5cm}{\centering PP warm-up, steady, \\ cool-down, and sync. \\state transition}}
\label{eq:wndcnt}
\end{equation}
\end{figure*}


\section{\sysname: Parallelism-driven Reconfiguration}

\begin{figure}[h!]
    \centering
    \includegraphics[width=1\linewidth]{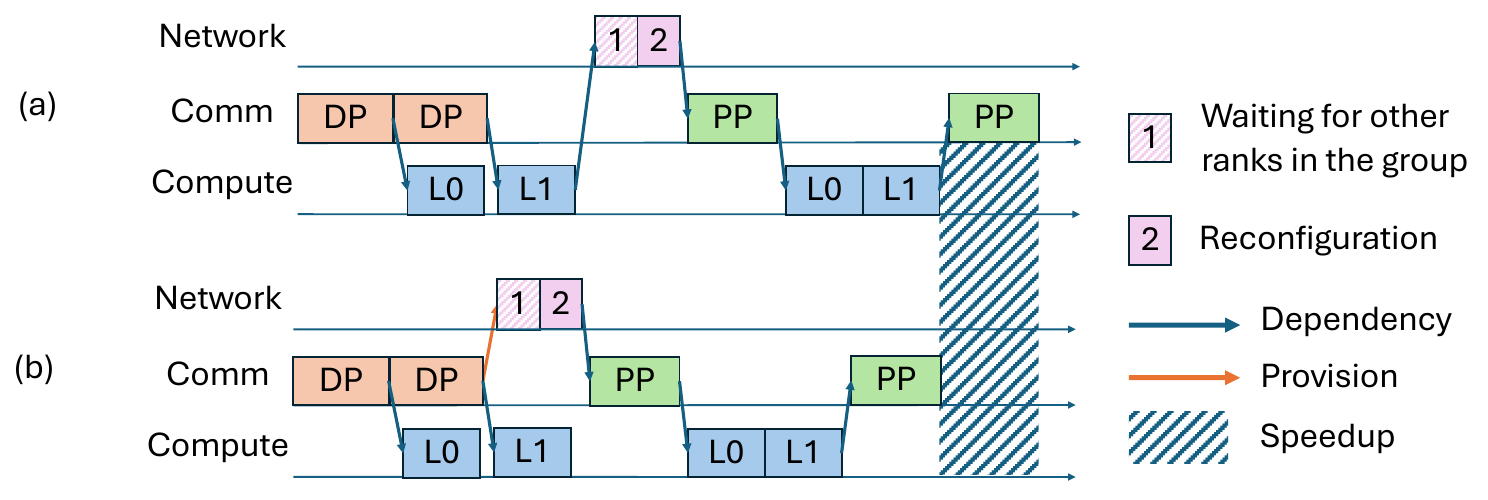}
    \caption{Reconfiguration during the warm-up stage of rank 0 and 4, (a) without provisioning, (b) with provisioning.}
    \label{fig:demo}
\end{figure}


We summarize several objectives for performing in-job topology reconfiguration: \textbf{Objective 1:} reconfiguring for new traffic demand with minimal delay; 
\textbf{Objective 2:} reducing reconfiguration frequency to increase circuit up-time and prevent blocking traffic; 
\textbf{Objective 3:} avoiding circuit conflicts (\eg new circuit interferes with ongoing traffic, demand from two ranks result in conflicting topology, \etc).

Meeting Objective (1) and (2) requires the control layer to accurately interpret the traffic patterns defined by the ML frameworks. Overlapping the reconfiguration delay with the traffic window is necessary to reduce overhead. To achieve this, the control layer can perform \emph{provisioning}, initiating the reconfiguration immediately once the previous communication kernel finishes (Fig. \ref{fig:demo}), if it identifies that the next communication belongs to a different parallelism. The control layer can also reduce reconfiguration frequency by only reconfiguring when there is a shift in parallelism. Fig. \ref{fig:config} shows all circuit configurations for the 3D-parallel workload.

To prevent circuit-level conflict with the workload DAG, a first-come-first-serve (FC-FS) scheduling policy is necessary in the control layer. A communication kernel which is issued first by the application should be served first within its communication group domain. The control layer should also prevent undesired control divergence across rails when collectives span multiple rails. Additionally, the reconfiguration can only be performed after the completion of the previous kernel, avoiding disrupting the ongoing traffic. 

To achieve those objectives, we believe \emph{the network control logic should be implemented in the application layer} where the parallelism-level hints and traffic patterns across iterations could be easily obtained. This design decision differs from previous control systems for reconfigurable networks in that they perform traffic and topology engineering on the packet or flow level in layer 2 and 3~\cite{cthrough,projector,farrington2010helios,amir2024shale,shoal, mellette2017rotornet, liang2024negotiator, ballani2020sirius}.

\subsection{Design Sketch}

\begin{figure}
    \centering
    \includegraphics[width=0.95\linewidth]{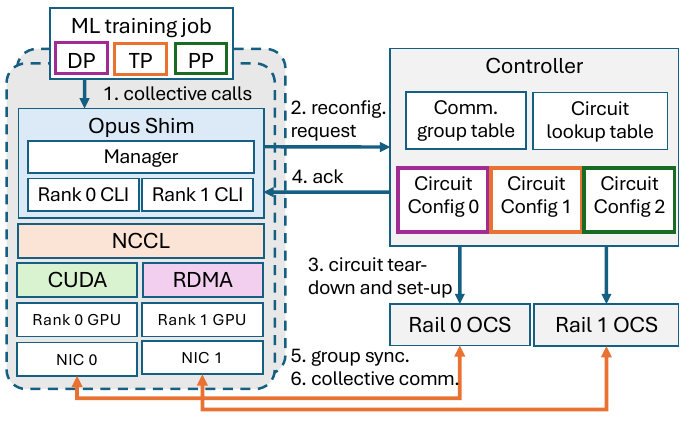}
    \caption{Opus control plane.}
    \label{fig:design}
\end{figure}

We present a high-level design sketch of the control plane, \sysname, for photonic rails, illustrated by Fig. \ref{fig:design}. 
\textbf{\sysname shim} runtime sits between the application and the collective communication layer in every scale-up domain.
By "intercepting" communication calls from the application layer, the shim gets the information of the traffic volume and group. The shim CLI profiles, interprets, and predicts traffic demand for individual ranks. The shim manager coalesces demand across rails  and issues reconfiguration requests. The manager also selects suitable networks based on traffic types, \eg the optical GPU-backend rail network for bulky data transfer, and high-bandwidth interconnect for intra-server traffic. \textbf{\sysname controller} orchestrates each rail's OCSes to perform reconfiguration upon receiving requests from all ranks of one communication group.
Together, the shim and the controller provide an illusion of an all-to-all GPU backend network, \emph{treating network connectivity as an allocatable resource}.

\myparab{During the first iteration of the training process,}
\sysname shim profiles the traffic pattern and issues reconfiguration request only if the demand matrix of the parallelism changes, reducing reconfiguration frequency. Upon receiving the requests,
\sysname controller populates per-communication-group metadata in a job-specific table, generates circuit configurations, and configures the OCSes. The shim waits for the controller's acknowledgment, and calls the collective communication library to execute the communication.

\myparab{During later iterations,} \sysname reduces reconfiguration overhead by \emph{provisioning}. The shim issues speculative requests based on the profiled schedules immediately after the completion of the previous traffic from a different communication group. The controller fetches the cached circuit configurations and programs the switches, allowing the subsequent communication being executed as soon as possible.
\subsection{\sysname Analysis}

\begin{figure}[h!]
    \centering
    \includegraphics[width=1\linewidth]{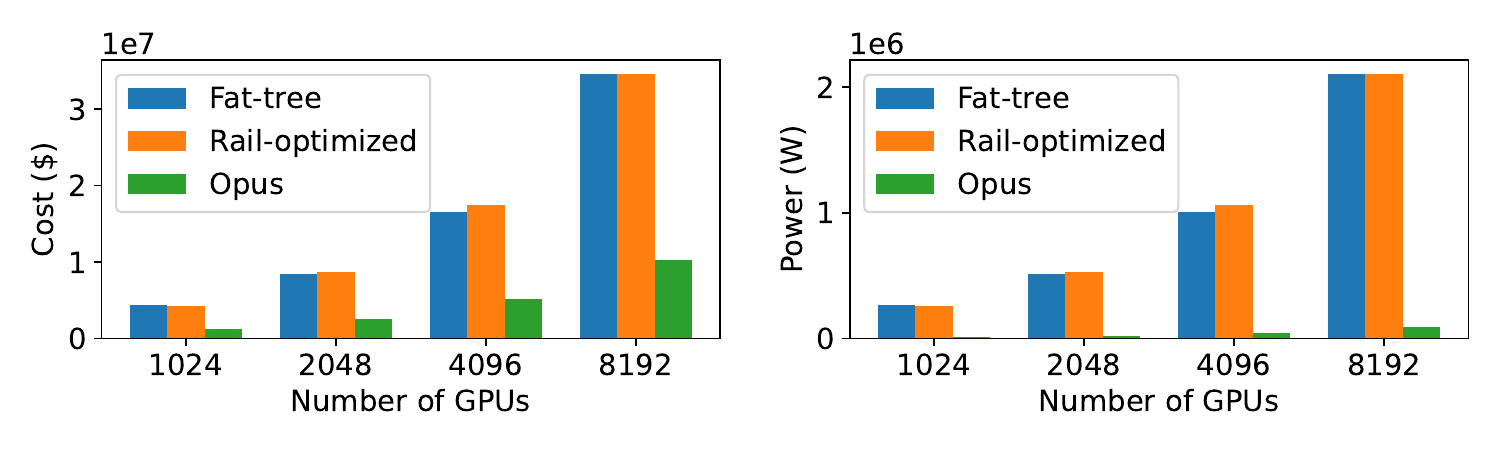}
    \caption{GPU-backend network cost and power comparison using DGX H200, 400Gbps transceivers and switches \cite{nvidia_connectx7_datasheet, fs_400gbase_xdr4_qsfpdd, fs_n9510_64d, polatis_series7000}, excluding fiber cable cost and power. }
    \label{fig:cost}
\end{figure}

\myparab{Scalability and energy efficiency.}
Using commodity switches, Opus GPU-backend network can scale up to 36K GPUs, and has a much lower cost and energy consumption compared to the state-of-the-art GPU networks. Tbl.~\ref{tab:opus_scalability_latency} shows the scalability--reconfiguration latency tradeoff for two scale-up domains using specifications from OCS vendors and prior works~\cite{telescent_products, polatis_series7000,  liao2025mfabric, calient2022sseries, mellette2017rotornet, lightmatter_passage, epiphotonics_products, sreenilayam2019fast, Coherent_OCS_2025}. We believe Piezo or 3D MEMS OCS is ideal for our design. To show \sysname's cost and power efficiency, we use the methodology from~\cite{wang2023topoopt, wang2024rail}. Opus saves cost by up to $70.5\%$ and power by up to $95.84\%$ (Fig. \ref{fig:cost}) thanks to the flat topology based on power-efficient OCSes and end-to-end optical data paths between GPUs.

\begin{table}[h!]
    \centering
    \footnotesize
    \begin{tabular}{lcccc}
        \hline
        \textbf{OCS Tech} & \makecell{\textbf{Reconfig.} \\ \textbf{time (ms)}} & \makecell{\textbf{Radix} \\ \textbf{(ports)}} & \makecell{\textbf{\# GPUs} \\ \textbf{(GB200)}} &  \makecell{\textbf{\# GPUs} \\ \textbf{(H200)}} \\
        \hline
        PLZT (EpiPhotonics)     & 0.00001  & 16   & 576  & 64   \\
        SiP (Lightmatter)       & 0.007    & 32  & 1152  & 128  \\
        RotorNet (InFocus)      & 0.01     & 128   & 4608  & 512  \\
        3D MEMS (Calient)       & 15       & 320  & 11520 & 1280 \\
        Piezo (Polatis)         & 25       & 576  & 20736 & 2304 \\
        Liquid crystal (Coherent) & 100    & 512  & 18432 & 2048 \\
        Robotic (Telescent)     & 120000   & 1008  & 36288 & 4032 \\
        \hline
    \end{tabular}
    \caption{Opus scalability-latency tradeoff. $\# \text{ GPUs} = \text{\# (GPUs in scale-up)} \times \text{ radix} / 2$ (using 2-port NIC configuration and bi-directional transceivers \cite{liu2023lightwave, jupiter-evolving}).}
    \label{tab:opus_scalability_latency}
\end{table}


\begin{figure}[h!]
    \centering
    \includegraphics[width=\linewidth]{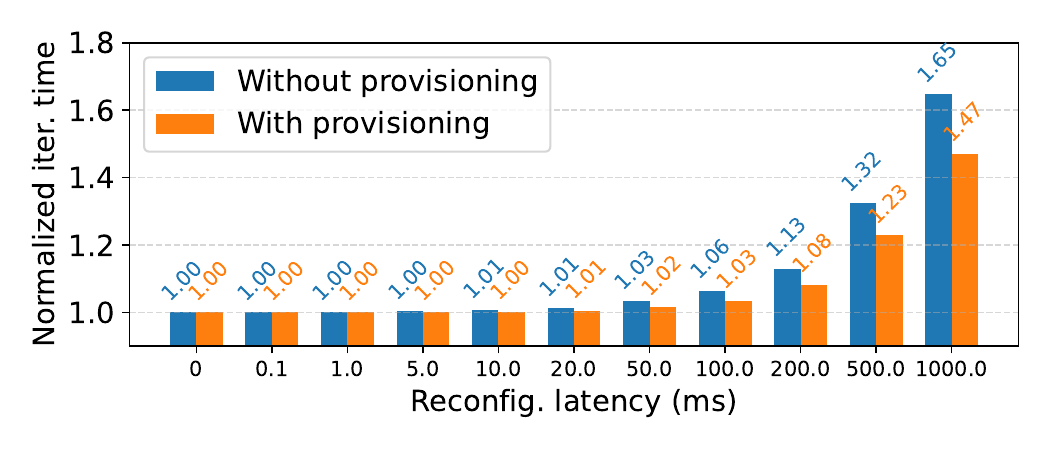}
    \caption{Iteration time for varying network reconfiguration delay (Llama3-8B with TorchTitan, TP=4, DP=PP=2).}
    \label{fig:itr}
    \vspace{-0.5cm}
\end{figure}

\myparab{Reconfiguration provisioning.}
We simulate the Llama3-8B training workload with optical rails using the trace in \S \ref{sec:trace}. Fig. \ref{fig:itr} shows the impact of reconfiguration delay on application when the rail reconfigures on the critical path (without provisioning), and proactively (with provisioning). The case of reconfiguration latency 0 stands for the baseline, a fully-connected network.  Since \sysname only reconfigures if the traffic pattern changes across parallelisms, the number of reconfiguration is kept small and the impact is minimal ($6.5\%$ iteration time increase with a 100ms switching delay over the baseline). The overhead could be minimized through provisioning. At 100ms switching delay,  the network only has a $3.5\%$ longer iteration than the baseline.

While our simulation assumes equal bandwidth between optical and electrical rails, optical interconnects have the potential to deliver significantly higher bandwidth, further reducing communication overhead in ML workloads \cite{cthrough, farrington2010helios}.   

\section{Discussion}
\myparab{Control plane and synchronization.}
Scheduling and time-synchronization have been major challenges in reconfigurable networks \cite{projector, liang2024negotiator, ballani2020sirius, amir2024shale}. Network elements need to be tightly synchronized to prevent packet loss. Centralized and distributed algorithms such as Edmonds and VLB \cite{cthrough, farrington2010helios, mellette2017rotornet} are used for packet delivery and load balancing. Our design achieves time-synchronization implicitly using the application-driven reconfiguration scheme and the collective barriers of collective communication libraries. The simple FC-FS scheduling is sufficient as the bandwidth resource is not shared across jobs, and there is a sequential ordering of traffic demands defined by the job framework.


\myparab{Supporting any communication patterns.}
\label{sec:anytraffic}
Optical rails form a physical ring connecting GPUs of the same rank in scale-out. This is suitable for \allreduce, \allgather, \reducescatter traffic, but is not optimal for certain traffic matrices that could not be easily implemented with a ring algorithm, such as \alltoall traffic in EP. Frequent switching among multiple communication groups (short \allreduce calls towards the end of a training iteration along PP and DP) is also challenging. A possible solution is to perform multi-hopping through connected GPUs in the same rail, and forward traffic through scale-up interconnect. The optical circuits could be configured to prioritize serving bottleneck traffic~\cite{liao2025mfabric}. Small, bursty traffic with high-incast could also be off-loaded to the host-based packet switched network~\cite{sato2023optical}.


\myparab{Reconfiguration granularity.}
Fine-grained reconfiguration in the optical rail network is required in hybrid parallelisms.
Though the parallelism dimension of a GPU's communication undergoes sequential orders, allowing us to define the time window for reconfiguration, traffic from different parallelisms may occur simultaneously in two groups of GPUs within one job. In the example of Fig. \ref{fig:config}(b), stage 2 ranks' DP \allgather and other stages' \sendrecv can be concurrent. 
Therefore, the circuit configurations should be conducted on the granularity of communication groups. 
Course-grained reconfiguration (directing all east-west links to north-south) would introduce conflicts with communication schedules of the ML frameworks. This demands flexible hardware switching for dynamic sets of ports (achievable by~\cite{polatis_series7000}) and per-rank control logic implementation. The flexibility is also important for multi-tenant environments.

\myparab{Opportunities.} Our reconfigurable rail design does not require modifications to the existing applications, as the control layer handles circuit allocation. However, the control layer has no leverage on the window sizes and collective orderings for hiding the reconfiguration delay.
Back-to-back or partially-overlapped traffic from two parallelism axes need to be spaced, resulting in bubbles and GPU idling.
We believe there are opportunities for further reducing the switching overhead by presenting the network resources (\ie circuit connectivity) as callable abstractions to the application, similar to the abstractions for compute acceleration (\eg \texttt{cuda.amp} for tensor cores in PyTorch \cite{pytorch_amp2025}). Utilizing the abstraction, the application, \eg PyTorch, could schedule computation kernels, communication collectives and the circuit allocation together, exploiting the overlapping possibilities between computation and network reconfiguration.

\section{Conclusion}
We propose a control layer to support ML workloads with hybrid parallelisms on reconfigurable optical rail-optimized networks. We show that performing circuit switching driven by parallelism shifts within a job can have $70.5\%$ cost saving and $95.84\%$ power reduction without significant impact on application performance, comparing to the fully-connected baselines. This is achieved through the tight integration between the application collective communication layer and the network controller, enabling fine-grained, parallelism-aware circuit provisioning and switching.

\bibliographystyle{ACM-Reference-Format} 
\bibliography{hotnets25-template}


\begin{thebibliography}{77}


\ifx \showCODEN    \undefined \def \showCODEN     #1{\unskip}     \fi
\ifx \showDOI      \undefined \def \showDOI       #1{#1}\fi
\ifx \showISBNx    \undefined \def \showISBNx     #1{\unskip}     \fi
\ifx \showISBNxiii \undefined \def \showISBNxiii  #1{\unskip}     \fi
\ifx \showISSN     \undefined \def \showISSN      #1{\unskip}     \fi
\ifx \showLCCN     \undefined \def \showLCCN      #1{\unskip}     \fi
\ifx \shownote     \undefined \def \shownote      #1{#1}          \fi
\ifx \showarticletitle \undefined \def \showarticletitle #1{#1}   \fi
\ifx \showURL      \undefined \def \showURL       {\relax}        \fi
\providecommand\bibfield[2]{#2}
\providecommand\bibinfo[2]{#2}
\providecommand\natexlab[1]{#1}
\providecommand\showeprint[2][]{arXiv:#2}

\bibitem[Agarwal et~al\mbox{.}(2024)]%
        {agarwal2024harmony}
\bibfield{author}{\bibinfo{person}{Saksham Agarwal}, \bibinfo{person}{Qizhe Cai}, \bibinfo{person}{Rachit Agarwal}, \bibinfo{person}{David Shmoys}, {and} \bibinfo{person}{Amin Vahdat}.} \bibinfo{year}{2024}\natexlab{}.
\newblock \showarticletitle{Harmony: A congestion-free datacenter architecture}. In \bibinfo{booktitle}{\emph{21st USENIX Symposium on Networked Systems Design and Implementation (NSDI 24)}}. \bibinfo{pages}{329--343}.
\newblock


\bibitem[Amir et~al\mbox{.}(2024)]%
        {amir2024shale}
\bibfield{author}{\bibinfo{person}{Daniel Amir}, \bibinfo{person}{Nitika Saran}, \bibinfo{person}{Tegan Wilson}, \bibinfo{person}{Robert Kleinberg}, \bibinfo{person}{Vishal Shrivastav}, {and} \bibinfo{person}{Hakim Weatherspoon}.} \bibinfo{year}{2024}\natexlab{}.
\newblock \showarticletitle{Shale: A practical, scalable oblivious reconfigurable network}. In \bibinfo{booktitle}{\emph{Proceedings of the ACM SIGCOMM 2024 Conference}}. \bibinfo{pages}{449--464}.
\newblock


\bibitem[Ballani et~al\mbox{.}(2020)]%
        {ballani2020sirius}
\bibfield{author}{\bibinfo{person}{Hitesh Ballani}, \bibinfo{person}{Paolo Costa}, \bibinfo{person}{Raphael Behrendt}, \bibinfo{person}{Daniel Cletheroe}, \bibinfo{person}{Istvan Haller}, \bibinfo{person}{Krzysztof Jozwik}, \bibinfo{person}{Fotini Karinou}, \bibinfo{person}{Sophie Lange}, \bibinfo{person}{Kai Shi}, \bibinfo{person}{Benn Thomsen}, {et~al\mbox{.}}} \bibinfo{year}{2020}\natexlab{}.
\newblock \showarticletitle{Sirius: A flat datacenter network with nanosecond optical switching}. In \bibinfo{booktitle}{\emph{Proceedings of the Annual conference of the ACM Special Interest Group on Data Communication on the applications, technologies, architectures, and protocols for computer communication}}. \bibinfo{pages}{782--797}.
\newblock


\bibitem[Besta and Hoefler(2014)]%
        {slimfly}
\bibfield{author}{\bibinfo{person}{Maciej Besta} {and} \bibinfo{person}{Torsten Hoefler}.} \bibinfo{year}{2014}\natexlab{}.
\newblock \showarticletitle{Slim Fly: A Cost Effective Low-Diameter Network Topology}. In \bibinfo{booktitle}{\emph{SC '14: Proceedings of the International Conference for High Performance Computing, Networking, Storage and Analysis}}. \bibinfo{pages}{348--359}.
\newblock
\urldef\tempurl%
\url{https://doi.org/10.1109/SC.2014.34}
\showDOI{\tempurl}


\bibitem[{Broadcom Inc.}(2025a)]%
        {broadcom_bcm78909}
\bibfield{author}{\bibinfo{person}{{Broadcom Inc.}}} \bibinfo{year}{2025}\natexlab{a}.
\newblock \bibinfo{title}{BCM78909 51.2‑Tb/s Multilayer Co‑Packaged Optics Switch}.
\newblock \bibinfo{howpublished}{Online; accessed July 5, 2025}.
\newblock
\urldef\tempurl%
\url{https://www.broadcom.com/products/fiber-optic-modules-components/co-packaged-optics/switches/bcm78909}
\showURL{%
\tempurl}
\newblock
\shownote{A high‑radix, high‑bandwidth CPO switch supporting up to 64×800GbE or 128×400GbE.}.


\bibitem[{Broadcom Inc.}(2025b)]%
        {broadcom_cpo}
\bibfield{author}{\bibinfo{person}{{Broadcom Inc.}}} \bibinfo{year}{2025}\natexlab{b}.
\newblock \bibinfo{title}{Co‑Packaged Optics (CPO)}.
\newblock \bibinfo{howpublished}{\url{https://www.broadcom.com/info/optics/cpo}}.
\newblock
\newblock
\shownote{Accessed: 2025-07-03}.


\bibitem[{Broadcom Inc.\, Optical Systems Division}(2021)]%
        {broadcom_scip_2021}
\bibfield{author}{\bibinfo{person}{{Broadcom Inc.\, Optical Systems Division}}.} \bibinfo{year}{2021}\natexlab{}.
\newblock \bibinfo{booktitle}{\emph{SiPh Chiplets In Package (SCIP)}}.
\newblock \bibinfo{type}{Technical Report}. \bibinfo{institution}{Broadcom Inc.}, \bibinfo{address}{Irvine, CA, USA}.
\newblock
\urldef\tempurl%
\url{https://docs.broadcom.com/doc/siph-chiplets-in-package-scip}
\showURL{%
\tempurl}
\newblock
\shownote{OSD CPO SCIP\_20211106 V5}.


\bibitem[{CALIENT Technologies, Inc.}(2022)]%
        {calient2022sseries}
\bibfield{author}{\bibinfo{person}{{CALIENT Technologies, Inc.}}} \bibinfo{year}{2022}\natexlab{}.
\newblock \bibinfo{title}{Calient’s Optical Circuit Switch (S‑Series) Datasheet}.
\newblock \bibinfo{howpublished}{\url{https://www.calient.net/wp-content/uploads/2022/06/Datasheet_Calients-Optical-Circuit-Switches.pdf}}.
\newblock
\newblock
\shownote{Accessed: 2025-07-03}.


\bibitem[Chen et~al\mbox{.}(2017)]%
        {chen2017enabling}
\bibfield{author}{\bibinfo{person}{Li Chen}, \bibinfo{person}{Kai Chen}, \bibinfo{person}{Zhonghua Zhu}, \bibinfo{person}{Minlan Yu}, \bibinfo{person}{George Porter}, \bibinfo{person}{Chunming Qiao}, {and} \bibinfo{person}{Shan Zhong}.} \bibinfo{year}{2017}\natexlab{}.
\newblock \showarticletitle{Enabling $\{$Wide-Spread$\}$ Communications on Optical Fabric with $\{$MegaSwitch$\}$}. In \bibinfo{booktitle}{\emph{14th USENIX Symposium on Networked Systems Design and Implementation (NSDI 17)}}. \bibinfo{pages}{577--593}.
\newblock


\bibitem[Chu et~al\mbox{.}(2025)]%
        {chu2025scaling}
\bibfield{author}{\bibinfo{person}{Weiwei Chu}, \bibinfo{person}{Xinfeng Xie}, \bibinfo{person}{Jiecao Yu}, \bibinfo{person}{Jie Wang}, \bibinfo{person}{Amar Phanishayee}, \bibinfo{person}{Chunqiang Tang}, \bibinfo{person}{Yuchen Hao}, \bibinfo{person}{Jianyu Huang}, \bibinfo{person}{Mustafa Ozdal}, \bibinfo{person}{Jun Wang}, {et~al\mbox{.}}} \bibinfo{year}{2025}\natexlab{}.
\newblock \showarticletitle{Scaling Llama 3 Training with Efficient Parallelism Strategies}. In \bibinfo{booktitle}{\emph{Proceedings of the 52nd Annual International Symposium on Computer Architecture}}. \bibinfo{pages}{1703--1716}.
\newblock


\bibitem[{Coherent Corp.}(2025)]%
        {Coherent_OCS_2025}
\bibfield{author}{\bibinfo{person}{{Coherent Corp.}}} \bibinfo{year}{2025}\natexlab{}.
\newblock \bibinfo{title}{Optical Circuit Switch (OCS)}.
\newblock \bibinfo{howpublished}{\url{https://www.coherent.com/networking/optical-circuit-switch}}.
\newblock
\newblock
\shownote{Accessed: 2025-07-10; Based on press release published March 25,2024; Coherent’s liquid‑crystal‑based OCS architecture supports up to 300×300 ports and is optimized for AI/ML data center fabrics}.


\bibitem[{EpiPhotonics Corp.}(2025)]%
        {epiphotonics_products}
\bibfield{author}{\bibinfo{person}{{EpiPhotonics Corp.}}} \bibinfo{year}{2025}\natexlab{}.
\newblock \bibinfo{title}{Products}.
\newblock \bibinfo{howpublished}{\url{http://epiphotonics.com/products.html}}.
\newblock
\newblock
\shownote{Accessed: 2025-07-03}.


\bibitem[Farrington et~al\mbox{.}(2010)]%
        {farrington2010helios}
\bibfield{author}{\bibinfo{person}{Nathan Farrington}, \bibinfo{person}{George Porter}, \bibinfo{person}{Sivasankar Radhakrishnan}, \bibinfo{person}{Hamid~Hajabdolali Bazzaz}, \bibinfo{person}{Vikram Subramanya}, \bibinfo{person}{Yeshaiahu Fainman}, \bibinfo{person}{George Papen}, {and} \bibinfo{person}{Amin Vahdat}.} \bibinfo{year}{2010}\natexlab{}.
\newblock \showarticletitle{Helios: a hybrid electrical/optical switch architecture for modular data centers}. In \bibinfo{booktitle}{\emph{Proceedings of the ACM SIGCOMM 2010 Conference}}. \bibinfo{pages}{339--350}.
\newblock


\bibitem[Fedus et~al\mbox{.}(2022)]%
        {fedus2022switch}
\bibfield{author}{\bibinfo{person}{William Fedus}, \bibinfo{person}{Barret Zoph}, {and} \bibinfo{person}{Noam Shazeer}.} \bibinfo{year}{2022}\natexlab{}.
\newblock \showarticletitle{Switch transformers: Scaling to trillion parameter models with simple and efficient sparsity}.
\newblock \bibinfo{journal}{\emph{Journal of Machine Learning Research}} \bibinfo{volume}{23}, \bibinfo{number}{120} (\bibinfo{year}{2022}), \bibinfo{pages}{1--39}.
\newblock


\bibitem[{FS.COM}(nda)]%
        {fs_400gbase_xdr4_qsfpdd}
\bibfield{author}{\bibinfo{person}{{FS.COM}}.} \bibinfo{year}{n.d.}\natexlab{a}.
\newblock \bibinfo{title}{{Cisco Compatible 400GBASE‑XDR4 QSFP‑DD PAM4 1310nm 2km Module}}.
\newblock \bibinfo{howpublished}{\url{https://www.fs.com/products/110530.html?attribute=94270&id=4477813}}.
\newblock
\newblock
\shownote{Accessed: 2025-07-02}.


\bibitem[{FS.COM}(ndb)]%
        {fs_n9510_64d}
\bibfield{author}{\bibinfo{person}{{FS.COM}}.} \bibinfo{year}{n.d.}\natexlab{b}.
\newblock \bibinfo{title}{{N9510‑64D 64‑Port Ethernet L3 Data Center Switch (Broadcom Tomahawk‑4, 64×400GbE)}}.
\newblock \bibinfo{howpublished}{\url{https://www.fs.com/products/149853.html}}.
\newblock
\newblock
\shownote{Accessed: 2025-07-02}.


\bibitem[Gangidi et~al\mbox{.}(2024)]%
        {gangidi2024rdma}
\bibfield{author}{\bibinfo{person}{Adithya Gangidi}, \bibinfo{person}{Rui Miao}, \bibinfo{person}{Shengbao Zheng}, \bibinfo{person}{Sai~Jayesh Bondu}, \bibinfo{person}{Guilherme Goes}, \bibinfo{person}{Hany Morsy}, \bibinfo{person}{Rohit Puri}, \bibinfo{person}{Mohammad Riftadi}, \bibinfo{person}{Ashmitha~Jeevaraj Shetty}, \bibinfo{person}{Jingyi Yang}, {et~al\mbox{.}}} \bibinfo{year}{2024}\natexlab{}.
\newblock \showarticletitle{Rdma over ethernet for distributed training at meta scale}. In \bibinfo{booktitle}{\emph{Proceedings of the ACM SIGCOMM 2024 Conference}}. \bibinfo{pages}{57--70}.
\newblock


\bibitem[Gherghescu et~al\mbox{.}(2024)]%
        {gherghescu2024ve}
\bibfield{author}{\bibinfo{person}{Alexandru~M Gherghescu}, \bibinfo{person}{Vlad-Andrei B{\u{a}}doiu}, \bibinfo{person}{Alexandru Agache}, \bibinfo{person}{Mihai-Valentin Dumitru}, \bibinfo{person}{Iuliu Vasilescu}, \bibinfo{person}{Radu Mantu}, {and} \bibinfo{person}{Costin Raiciu}.} \bibinfo{year}{2024}\natexlab{}.
\newblock \showarticletitle{I've Got 99 Problems But FLOPS Ain't One}. In \bibinfo{booktitle}{\emph{Proceedings of the 23rd ACM Workshop on Hot Topics in Networks}}. \bibinfo{pages}{195--204}.
\newblock


\bibitem[Ghobadi et~al\mbox{.}(2016)]%
        {projector}
\bibfield{author}{\bibinfo{person}{Monia Ghobadi}, \bibinfo{person}{Ratul Mahajan}, \bibinfo{person}{Amar Phanishayee}, \bibinfo{person}{Nikhil Devanur}, \bibinfo{person}{Janardhan Kulkarni}, \bibinfo{person}{Gireeja Ranade}, \bibinfo{person}{Pierre-Alexandre Blanche}, \bibinfo{person}{Houman Rastegarfar}, \bibinfo{person}{Madeleine Glick}, {and} \bibinfo{person}{Daniel Kilper}.} \bibinfo{year}{2016}\natexlab{}.
\newblock \showarticletitle{ProjecToR: Agile Reconfigurable Data Center Interconnect}. In \bibinfo{booktitle}{\emph{Proceedings of the 2016 ACM SIGCOMM Conference}} (Florianopolis, Brazil) \emph{(\bibinfo{series}{SIGCOMM '16})}. \bibinfo{publisher}{Association for Computing Machinery}, \bibinfo{address}{New York, NY, USA}, \bibinfo{pages}{216–229}.
\newblock
\showISBNx{9781450341936}
\urldef\tempurl%
\url{https://doi.org/10.1145/2934872.2934911}
\showDOI{\tempurl}


\bibitem[Grattafiori et~al\mbox{.}(2024)]%
        {grattafiori2024llama}
\bibfield{author}{\bibinfo{person}{Aaron Grattafiori}, \bibinfo{person}{Abhimanyu Dubey}, \bibinfo{person}{Abhinav Jauhri}, \bibinfo{person}{Abhinav Pandey}, \bibinfo{person}{Abhishek Kadian}, \bibinfo{person}{Ahmad Al-Dahle}, \bibinfo{person}{Aiesha Letman}, \bibinfo{person}{Akhil Mathur}, \bibinfo{person}{Alan Schelten}, \bibinfo{person}{Alex Vaughan}, {et~al\mbox{.}}} \bibinfo{year}{2024}\natexlab{}.
\newblock \showarticletitle{The llama 3 herd of models}.
\newblock \bibinfo{journal}{\emph{arXiv preprint arXiv:2407.21783}} (\bibinfo{year}{2024}).
\newblock


\bibitem[Greenberg et~al\mbox{.}(2009)]%
        {vl2}
\bibfield{author}{\bibinfo{person}{Albert Greenberg}, \bibinfo{person}{James~R. Hamilton}, \bibinfo{person}{Navendu Jain}, \bibinfo{person}{Srikanth Kandula}, \bibinfo{person}{Changhoon Kim}, \bibinfo{person}{Parantap Lahiri}, \bibinfo{person}{David~A. Maltz}, \bibinfo{person}{Parveen Patel}, {and} \bibinfo{person}{Sudipta Sengupta}.} \bibinfo{year}{2009}\natexlab{}.
\newblock \showarticletitle{VL2: A Scalable and Flexible Data Center Network}. In \bibinfo{booktitle}{\emph{Proceedings of the ACM SIGCOMM 2009 Conference on Data Communication}} (Barcelona, Spain) \emph{(\bibinfo{series}{SIGCOMM '09})}. \bibinfo{publisher}{Association for Computing Machinery}, \bibinfo{address}{New York, NY, USA}, \bibinfo{pages}{51–62}.
\newblock
\showISBNx{9781605585949}
\urldef\tempurl%
\url{https://doi.org/10.1145/1592568.1592576}
\showDOI{\tempurl}


\bibitem[Hamedazimi et~al\mbox{.}(2014)]%
        {firefly}
\bibfield{author}{\bibinfo{person}{Navid Hamedazimi}, \bibinfo{person}{Zafar Qazi}, \bibinfo{person}{Himanshu Gupta}, \bibinfo{person}{Vyas Sekar}, \bibinfo{person}{Samir~R. Das}, \bibinfo{person}{Jon~P. Longtin}, \bibinfo{person}{Himanshu Shah}, {and} \bibinfo{person}{Ashish Tanwer}.} \bibinfo{year}{2014}\natexlab{}.
\newblock \showarticletitle{FireFly: A Reconfigurable Wireless Data Center Fabric Using Free-Space Optics}. In \bibinfo{booktitle}{\emph{Proceedings of the 2014 ACM Conference on SIGCOMM}} (Chicago, Illinois, USA) \emph{(\bibinfo{series}{SIGCOMM '14})}. \bibinfo{publisher}{Association for Computing Machinery}, \bibinfo{address}{New York, NY, USA}, \bibinfo{pages}{319–330}.
\newblock
\showISBNx{9781450328364}
\urldef\tempurl%
\url{https://doi.org/10.1145/2619239.2626328}
\showDOI{\tempurl}


\bibitem[Harsh et~al\mbox{.}(2020)]%
        {harsh2020spineless}
\bibfield{author}{\bibinfo{person}{Vipul Harsh}, \bibinfo{person}{Sangeetha~Abdu Jyothi}, {and} \bibinfo{person}{P~Brighten Godfrey}.} \bibinfo{year}{2020}\natexlab{}.
\newblock \showarticletitle{Spineless data centers}. In \bibinfo{booktitle}{\emph{Proceedings of the 19th ACM Workshop on Hot Topics in Networks}}. \bibinfo{pages}{67--73}.
\newblock


\bibitem[{Hewlett Packard Enterprise}(2021)]%
        {hpe_cray_ex}
\bibfield{author}{\bibinfo{person}{{Hewlett Packard Enterprise}}.} \bibinfo{year}{2021}\natexlab{}.
\newblock \bibinfo{title}{HPE Cray EX Supercomputer Overview}.
\newblock \bibinfo{howpublished}{\url{https://www.hpe.com/psnow/doc/a50002546enw}}.
\newblock
\newblock
\shownote{Accessed: 2025-07-09}.


\bibitem[Jouppi et~al\mbox{.}(2023)]%
        {jouppi2023tpu}
\bibfield{author}{\bibinfo{person}{Norm Jouppi}, \bibinfo{person}{George Kurian}, \bibinfo{person}{Sheng Li}, \bibinfo{person}{Peter Ma}, \bibinfo{person}{Rahul Nagarajan}, \bibinfo{person}{Lifeng Nai}, \bibinfo{person}{Nishant Patil}, \bibinfo{person}{Suvinay Subramanian}, \bibinfo{person}{Andy Swing}, \bibinfo{person}{Brian Towles}, {et~al\mbox{.}}} \bibinfo{year}{2023}\natexlab{}.
\newblock \showarticletitle{Tpu v4: An optically reconfigurable supercomputer for machine learning with hardware support for embeddings}. In \bibinfo{booktitle}{\emph{Proceedings of the 50th annual international symposium on computer architecture}}. \bibinfo{pages}{1--14}.
\newblock


\bibitem[Khani et~al\mbox{.}({[n.\,d.]})]%
        {sipml}
\bibfield{author}{\bibinfo{person}{Mehrdad Khani}, \bibinfo{person}{Manya Ghobadi}, \bibinfo{person}{Mohammad Alizadeh}, \bibinfo{person}{Ziyi Zhu}, \bibinfo{person}{Madeleine Glick}, \bibinfo{person}{Keren Bergman}, \bibinfo{person}{Amin Vahdat}, \bibinfo{person}{Benjamin Klenk}, {and} \bibinfo{person}{Eiman Ebrahimi}.} \bibinfo{year}{[n.\,d.]}\natexlab{}.
\newblock \showarticletitle{SiP-ML: High-Bandwidth Optical Network Interconnects for Machine Learning Training}. In \bibinfo{booktitle}{\emph{Proceedings of the 2021 ACM SIGCOMM 2021 Conference}}.
\newblock


\bibitem[Korthikanti et~al\mbox{.}(2023)]%
        {korthikanti2023reducing}
\bibfield{author}{\bibinfo{person}{Vijay~Anand Korthikanti}, \bibinfo{person}{Jared Casper}, \bibinfo{person}{Sangkug Lym}, \bibinfo{person}{Lawrence McAfee}, \bibinfo{person}{Michael Andersch}, \bibinfo{person}{Mohammad Shoeybi}, {and} \bibinfo{person}{Bryan Catanzaro}.} \bibinfo{year}{2023}\natexlab{}.
\newblock \showarticletitle{Reducing activation recomputation in large transformer models}.
\newblock \bibinfo{journal}{\emph{Proceedings of Machine Learning and Systems}}  \bibinfo{volume}{5} (\bibinfo{year}{2023}), \bibinfo{pages}{341--353}.
\newblock


\bibitem[Kumar et~al\mbox{.}(2024)]%
        {lumorph}
\bibfield{author}{\bibinfo{person}{Abhishek~Vijaya Kumar}, \bibinfo{person}{Arjun Devraj}, \bibinfo{person}{Darius Bunandar}, {and} \bibinfo{person}{Rachee Singh}.} \bibinfo{year}{2024}\natexlab{}.
\newblock \showarticletitle{A case for server-scale photonic connectivity}. In \bibinfo{booktitle}{\emph{Proceedings of the 23rd ACM Workshop on Hot Topics in Networks}} (Irvine, CA, USA) \emph{(\bibinfo{series}{HotNets '24})}. \bibinfo{publisher}{Association for Computing Machinery}, \bibinfo{address}{New York, NY, USA}, \bibinfo{pages}{290–299}.
\newblock
\showISBNx{9798400712722}
\urldef\tempurl%
\url{https://doi.org/10.1145/3696348.3696856}
\showDOI{\tempurl}


\bibitem[Kumar et~al\mbox{.}(2025)]%
        {lumion}
\bibfield{author}{\bibinfo{person}{Abhishek~Vijaya Kumar}, \bibinfo{person}{Eric Ding}, \bibinfo{person}{Arjun Devraj}, \bibinfo{person}{Darius Bunandar}, {and} \bibinfo{person}{Rachee Singh}.} \bibinfo{year}{2025}\natexlab{}.
\newblock \bibinfo{title}{LUMION: Fast Fault Recovery for ML Jobs Using Programmable Optical Fabrics}.
\newblock
\newblock
\showeprint[arxiv]{2505.23105}~[cs.LG]
\urldef\tempurl%
\url{https://arxiv.org/abs/2505.23105}
\showURL{%
\tempurl}


\bibitem[Liang et~al\mbox{.}(2024b)]%
        {liang2024negotiator}
\bibfield{author}{\bibinfo{person}{Cong Liang}, \bibinfo{person}{Xiangli Song}, \bibinfo{person}{Jing Cheng}, \bibinfo{person}{Mowei Wang}, \bibinfo{person}{Yashe Liu}, \bibinfo{person}{Zhenhua Liu}, \bibinfo{person}{Shizhen Zhao}, {and} \bibinfo{person}{Yong Cui}.} \bibinfo{year}{2024}\natexlab{b}.
\newblock \showarticletitle{NegotiaToR: Towards A Simple Yet Effective On-demand Reconfigurable Datacenter Network}. In \bibinfo{booktitle}{\emph{Proceedings of the ACM SIGCOMM 2024 Conference}}. \bibinfo{pages}{415--432}.
\newblock


\bibitem[Liang et~al\mbox{.}(2024a)]%
        {liang2024torchtitan}
\bibfield{author}{\bibinfo{person}{Wanchao Liang}, \bibinfo{person}{Tianyu Liu}, \bibinfo{person}{Less Wright}, \bibinfo{person}{Will Constable}, \bibinfo{person}{Andrew Gu}, \bibinfo{person}{Chien-Chin Huang}, \bibinfo{person}{Iris Zhang}, \bibinfo{person}{Wei Feng}, \bibinfo{person}{Howard Huang}, \bibinfo{person}{Junjie Wang}, {et~al\mbox{.}}} \bibinfo{year}{2024}\natexlab{a}.
\newblock \showarticletitle{TorchTitan: One-stop PyTorch native solution for production ready LLM pre-training}.
\newblock \bibinfo{journal}{\emph{arXiv preprint arXiv:2410.06511}} (\bibinfo{year}{2024}).
\newblock


\bibitem[Liao et~al\mbox{.}(2025)]%
        {liao2025mfabric}
\bibfield{author}{\bibinfo{person}{Xudong Liao}, \bibinfo{person}{Yijun Sun}, \bibinfo{person}{Han Tian}, \bibinfo{person}{Xinchen Wan}, \bibinfo{person}{Yilun Jin}, \bibinfo{person}{Zilong Wang}, \bibinfo{person}{Zhenghang Ren}, \bibinfo{person}{Xinyang Huang}, \bibinfo{person}{Wenxue Li}, \bibinfo{person}{Kin~Fai Tse}, {et~al\mbox{.}}} \bibinfo{year}{2025}\natexlab{}.
\newblock \showarticletitle{mFabric: An Efficient and Scalable Fabric for Mixture-of-Experts Training}.
\newblock \bibinfo{journal}{\emph{arXiv preprint arXiv:2501.03905}} (\bibinfo{year}{2025}).
\newblock


\bibitem[{Lightmatter, Inc.}(2025)]%
        {lightmatter_passage}
\bibfield{author}{\bibinfo{person}{{Lightmatter, Inc.}}} \bibinfo{year}{2025}\natexlab{}.
\newblock \bibinfo{title}{Passage Technology}.
\newblock \bibinfo{howpublished}{\url{https://lightmatter.co/products/passage/}}.
\newblock
\newblock
\shownote{Accessed: 2025-07-03}.


\bibitem[Liu et~al\mbox{.}(2024)]%
        {liu2024deepseek}
\bibfield{author}{\bibinfo{person}{Aixin Liu}, \bibinfo{person}{Bei Feng}, \bibinfo{person}{Bing Xue}, \bibinfo{person}{Bingxuan Wang}, \bibinfo{person}{Bochao Wu}, \bibinfo{person}{Chengda Lu}, \bibinfo{person}{Chenggang Zhao}, \bibinfo{person}{Chengqi Deng}, \bibinfo{person}{Chenyu Zhang}, \bibinfo{person}{Chong Ruan}, {et~al\mbox{.}}} \bibinfo{year}{2024}\natexlab{}.
\newblock \showarticletitle{Deepseek-v3 technical report}.
\newblock \bibinfo{journal}{\emph{arXiv preprint arXiv:2412.19437}} (\bibinfo{year}{2024}).
\newblock


\bibitem[Liu et~al\mbox{.}(2023a)]%
        {liu2023lightwave}
\bibfield{author}{\bibinfo{person}{Hong Liu}, \bibinfo{person}{Ryohei Urata}, \bibinfo{person}{Kevin Yasumura}, \bibinfo{person}{Xiang Zhou}, \bibinfo{person}{Roy Bannon}, \bibinfo{person}{Jill Berger}, \bibinfo{person}{Pedram Dashti}, \bibinfo{person}{Norm Jouppi}, \bibinfo{person}{Cedric Lam}, \bibinfo{person}{Sheng Li}, {et~al\mbox{.}}} \bibinfo{year}{2023}\natexlab{a}.
\newblock \showarticletitle{Lightwave fabrics: at-scale optical circuit switching for datacenter and machine learning systems}. In \bibinfo{booktitle}{\emph{Proceedings of the ACM SIGCOMM 2023 Conference}}. \bibinfo{pages}{499--515}.
\newblock


\bibitem[Liu et~al\mbox{.}(2023b)]%
        {liu2023ring}
\bibfield{author}{\bibinfo{person}{Hao Liu}, \bibinfo{person}{Matei Zaharia}, {and} \bibinfo{person}{Pieter Abbeel}.} \bibinfo{year}{2023}\natexlab{b}.
\newblock \showarticletitle{Ring attention with blockwise transformers for near-infinite context}.
\newblock \bibinfo{journal}{\emph{arXiv preprint arXiv:2310.01889}} (\bibinfo{year}{2023}).
\newblock


\bibitem[{Lumentum Holdings Inc.}(2025)]%
        {Lumentum2025OCS}
\bibfield{author}{\bibinfo{person}{{Lumentum Holdings Inc.}}} \bibinfo{year}{2025}\natexlab{}.
\newblock \bibinfo{title}{{Lumentum Optical Circuit Switch to Improve Next‑Generation AI Data Center Scalability}}.
\newblock \bibinfo{howpublished}{\url{https://www.lumentum.com/en/media-room/news-releases/lumentum-optical-circuit-switch-improve-next-generation-ai-data-center}}.
\newblock
\newblock
\shownote{Accessed June 20, 2025}.


\bibitem[Mellette et~al\mbox{.}(2017)]%
        {mellette2017rotornet}
\bibfield{author}{\bibinfo{person}{William~M Mellette}, \bibinfo{person}{Rob McGuinness}, \bibinfo{person}{Arjun Roy}, \bibinfo{person}{Alex Forencich}, \bibinfo{person}{George Papen}, \bibinfo{person}{Alex~C Snoeren}, {and} \bibinfo{person}{George Porter}.} \bibinfo{year}{2017}\natexlab{}.
\newblock \showarticletitle{Rotornet: A scalable, low-complexity, optical datacenter network}. In \bibinfo{booktitle}{\emph{Proceedings of the Conference of the ACM Special Interest Group on Data Communication}}. \bibinfo{pages}{267--280}.
\newblock


\bibitem[Micikevicius et~al\mbox{.}(2017)]%
        {micikevicius2017mixed}
\bibfield{author}{\bibinfo{person}{Paulius Micikevicius}, \bibinfo{person}{Sharan Narang}, \bibinfo{person}{Jonah Alben}, \bibinfo{person}{Gregory Diamos}, \bibinfo{person}{Erich Elsen}, \bibinfo{person}{David Garcia}, \bibinfo{person}{Boris Ginsburg}, \bibinfo{person}{Michael Houston}, \bibinfo{person}{Oleksii Kuchaiev}, \bibinfo{person}{Ganesh Venkatesh}, {et~al\mbox{.}}} \bibinfo{year}{2017}\natexlab{}.
\newblock \showarticletitle{Mixed precision training}.
\newblock \bibinfo{journal}{\emph{arXiv preprint arXiv:1710.03740}} (\bibinfo{year}{2017}).
\newblock


\bibitem[{National Energy Research Scientific Computing Center (NERSC)}(2025)]%
        {nersc_perlmutter_architecture}
\bibfield{author}{\bibinfo{person}{{National Energy Research Scientific Computing Center (NERSC)}}.} \bibinfo{year}{2025}\natexlab{}.
\newblock \bibinfo{title}{{Perlmutter Architecture — NERSC Documentation}}.
\newblock \bibinfo{howpublished}{\url{https://docs.nersc.gov/systems/perlmutter/architecture/}}.
\newblock
\newblock
\shownote{Accessed: 2025-07-04}.


\bibitem[{NVIDIA}(2025)]%
        {nvidia2025llama31_405b_dgxc}
\bibfield{author}{\bibinfo{person}{{NVIDIA}}.} \bibinfo{year}{2025}\natexlab{}.
\newblock \bibinfo{title}{{Llama-3.1-405B DGXC Benchmarking Recipe}}.
\newblock \bibinfo{howpublished}{\url{https://catalog.ngc.nvidia.com/orgs/nvidia/teams/dgxc-benchmarking/resources/llama31-405b-dgxc-benchmarking-a}}.
\newblock
\newblock
\shownote{Version 24.11.1, modified January 29, 2025}.


\bibitem[{NVIDIA Corporation}(2020)]%
        {nvidia_nccl_communicators}
\bibfield{author}{\bibinfo{person}{{NVIDIA Corporation}}.} \bibinfo{year}{2020}\natexlab{}.
\newblock \bibinfo{booktitle}{\emph{{NVIDIA Collective Communication Library (NCCL)}: Creating a Communicator}}.
\newblock {NVIDIA}.
\newblock
\urldef\tempurl%
\url{https://docs.nvidia.com/deeplearning/nccl/user-guide/docs/usage/communicators.html}
\showURL{%
\tempurl}
\newblock
\shownote{Accessed July 6, 2025}.


\bibitem[{NVIDIA Corporation}(2022)]%
        {nvidia-pxn}
\bibfield{author}{\bibinfo{person}{{NVIDIA Corporation}}.} \bibinfo{year}{2022}\natexlab{}.
\newblock \bibinfo{title}{Doubling all‑to‑all Performance with NCCL 2.12: Introducing PXN (PCI X NVLink)}.
\newblock \bibinfo{howpublished}{NVIDIA Developer Blog}.
\newblock
\urldef\tempurl%
\url{https://developer.nvidia.com/blog/doubling-all2all-performance-with-nvidia-collective-communication-library-2-12/}
\showURL{%
\tempurl}
\newblock
\shownote{Describes PXN, which enables GPU‐to‐NIC communication via NVLink to optimize rail‐aligned collective performance}.


\bibitem[{NVIDIA Corporation}(2024)]%
        {nvidia_connectx7_datasheet}
\bibfield{author}{\bibinfo{person}{{NVIDIA Corporation}}.} \bibinfo{year}{2024}\natexlab{}.
\newblock \bibinfo{title}{{ConnectX‑7 400G Adapters Datasheet}}.
\newblock \bibinfo{howpublished}{\url{https://resources.nvidia.com/en-us-accelerated-networking-resource-library/connectx-7-datasheet}}.
\newblock
\newblock
\shownote{Accessed: 2025-07-02}.


\bibitem[{NVIDIA Corporation}(2025a)]%
        {nvidia_silicon_photonics}
\bibfield{author}{\bibinfo{person}{{NVIDIA Corporation}}.} \bibinfo{year}{2025}\natexlab{a}.
\newblock \bibinfo{title}{Co‑Packaged Silicon Photonics Networking Switches}.
\newblock \bibinfo{howpublished}{Online; accessed July 5,2025}.
\newblock
\urldef\tempurl%
\url{https://www.nvidia.com/en-us/networking/products/silicon-photonics/}
\showURL{%
\tempurl}
\newblock
\shownote{Describes NVIDIA’s co‑packaged optics (CPO) switches with integrated silicon photonics}.


\bibitem[{NVIDIA Corporation}(2025b)]%
        {nvidia_spectrum-x_2025}
\bibfield{author}{\bibinfo{person}{{NVIDIA Corporation}}.} \bibinfo{year}{2025}\natexlab{b}.
\newblock \bibinfo{booktitle}{\emph{NVIDIA Announces Spectrum‑X Photonics, Co‑Packaged Optics Networking Switches to Scale AI Factories to Millions of GPUs}}.
\newblock \bibinfo{type}{Press Release}. \bibinfo{institution}{NVIDIA Corporation}, \bibinfo{address}{Santa Clara, CA, USA}.
\newblock
\urldef\tempurl%
\url{https://nvidianews.nvidia.com/news/nvidia-spectrum-x-co-packaged-optics-networking-switches-ai-factories}
\showURL{%
\tempurl}
\newblock
\shownote{Unveiled at GTC 2025}.


\bibitem[{NVIDIA Corporation}(2025c)]%
        {nvidia_nccl}
\bibfield{author}{\bibinfo{person}{{NVIDIA Corporation}}.} \bibinfo{year}{2025}\natexlab{c}.
\newblock \bibinfo{booktitle}{\emph{{NVIDIA Collective Communications Library (NCCL)}}}.
\newblock NVIDIA Developer.
\newblock
\urldef\tempurl%
\url{https://developer.nvidia.com/nccl}
\showURL{%
\tempurl}
\newblock
\shownote{Version 2.x; MPI-compatible multi‑GPU / multi‑node collective communication library}.


\bibitem[{NVIDIA Corporation}(2025d)]%
        {nvidia2025dgx_h200}
\bibfield{author}{\bibinfo{person}{{NVIDIA Corporation}}.} \bibinfo{year}{2025}\natexlab{d}.
\newblock \bibinfo{booktitle}{\emph{NVIDIA DGX H200 Datasheet}}.
\newblock \bibinfo{type}{Datasheet}. \bibinfo{institution}{NVIDIA Corporation}, \bibinfo{address}{Santa Clara, CA}.
\newblock
\urldef\tempurl%
\url{https://resources.nvidia.com/en-us-dgx-systems/dgx-h200-datasheet}
\showURL{%
\tempurl}
\newblock
\shownote{Includes specifications of the DGX H200 system, featuring 8× H200 GPUs, dual Xeon Platinum 8480C CPUs, 2 TB system memory, 30 TB NVMe SSD, and full NVIDIA AI Enterprise software stack}.


\bibitem[{NVIDIA Corporation}(2025e)]%
        {nvidia_dgx_superpod_2025}
\bibfield{author}{\bibinfo{person}{{NVIDIA Corporation}}.} \bibinfo{year}{2025}\natexlab{e}.
\newblock \bibinfo{booktitle}{\emph{{NVIDIA DGX SuperPOD}}}.
\newblock NVIDIA.
\newblock
\urldef\tempurl%
\url{https://www.nvidia.com/en-us/data-center/dgx-superpod/}
\showURL{%
\tempurl}
\newblock
\shownote{Full-stack data center platform scaling to tens of thousands of GPUs; includes compute, networking, storage, and software}.


\bibitem[{NVIDIA Corporation}(2025f)]%
        {nvidia_hgx_platform_2025}
\bibfield{author}{\bibinfo{person}{{NVIDIA Corporation}}.} \bibinfo{year}{2025}\natexlab{f}.
\newblock \bibinfo{booktitle}{\emph{{NVIDIA HGX Platform}}}.
\newblock NVIDIA.
\newblock
\urldef\tempurl%
\url{https://www.nvidia.com/en-us/data-center/hgx/}
\showURL{%
\tempurl}
\newblock
\shownote{Reference architecture combining GPUs, NVLink/NVSwitch, networking, and AI/HPC software stack}.


\bibitem[{NVIDIA Corporation}(2025g)]%
        {nvidia-rail-optimize}
\bibfield{author}{\bibinfo{person}{{NVIDIA Corporation}}.} \bibinfo{year}{2025}\natexlab{g}.
\newblock \bibinfo{booktitle}{\emph{Rail Optimized Topology Validation}}.
\newblock NVIDIA Networking, Santa Clara, CA.
\newblock
\urldef\tempurl%
\url{https://docs.nvidia.com/networking/display/ibdiagnetusermanualv221/Rail+Optimized+Topology+Validation}
\showURL{%
\tempurl}
\newblock
\shownote{Part of the ibdiagnet InfiniBand Fabric Diagnostic Tool User Manual; describes cabling validation and compute-fabric alignment in DGX SuperPOD rail‑optimized fabrics}.


\bibitem[Pascanu et~al\mbox{.}(2013)]%
        {pascanu2013difficulty}
\bibfield{author}{\bibinfo{person}{Razvan Pascanu}, \bibinfo{person}{Tomas Mikolov}, {and} \bibinfo{person}{Yoshua Bengio}.} \bibinfo{year}{2013}\natexlab{}.
\newblock \showarticletitle{On the difficulty of training recurrent neural networks}. In \bibinfo{booktitle}{\emph{International conference on machine learning}}. Pmlr, \bibinfo{pages}{1310--1318}.
\newblock


\bibitem[{Polatis (a HUBER+SUHNER company)}(nd)]%
        {polatis_series7000}
\bibfield{author}{\bibinfo{person}{{Polatis (a HUBER+SUHNER company)}}.} \bibinfo{year}{n.d.}\natexlab{}.
\newblock \bibinfo{title}{Series 7000 — 384×384‑port Software‑Defined Optical Circuit Switch}.
\newblock \bibinfo{howpublished}{\url{https://www.polatis.com/series-7000-384x384-port-software-controlled-optical-circuit-switch-sdn-enabled.asp}}.
\newblock
\newblock
\shownote{Accessed: 2025-07-01}.


\bibitem[Poutievski et~al\mbox{.}(2022)]%
        {jupiter-evolving}
\bibfield{author}{\bibinfo{person}{Leon Poutievski}, \bibinfo{person}{Omid Mashayekhi}, \bibinfo{person}{Joon Ong}, \bibinfo{person}{Arjun Singh}, \bibinfo{person}{Mukarram Tariq}, \bibinfo{person}{Rui Wang}, \bibinfo{person}{Jianan Zhang}, \bibinfo{person}{Virginia Beauregard}, \bibinfo{person}{Patrick Conner}, \bibinfo{person}{Steve Gribble}, \bibinfo{person}{Rishi Kapoor}, \bibinfo{person}{Stephen Kratzer}, \bibinfo{person}{Nanfang Li}, \bibinfo{person}{Hong Liu}, \bibinfo{person}{Karthik Nagaraj}, \bibinfo{person}{Jason Ornstein}, \bibinfo{person}{Samir Sawhney}, \bibinfo{person}{Ryohei Urata}, \bibinfo{person}{Lorenzo Vicisano}, \bibinfo{person}{Kevin Yasumura}, \bibinfo{person}{Shidong Zhang}, \bibinfo{person}{Junlan Zhou}, {and} \bibinfo{person}{Amin Vahdat}.} \bibinfo{year}{2022}\natexlab{}.
\newblock \showarticletitle{{Jupiter Evolving: Transforming Google's Datacenter Network via Optical Circuit Switches and Software-Defined Networking}}. In \bibinfo{booktitle}{\emph{Proceedings of the ACM SIGCOMM 2022 Conference}} \emph{(\bibinfo{series}{SIGCOMM '22})}. \bibinfo{pages}{66–85}.
\newblock


\bibitem[{PyTorch Team}(2025)]%
        {pytorch_amp2025}
\bibfield{author}{\bibinfo{person}{{PyTorch Team}}.} \bibinfo{year}{2025}\natexlab{}.
\newblock \bibinfo{title}{Automatic Mixed Precision package (\texttt{torch.amp})}.
\newblock \bibinfo{howpublished}{\url{https://pytorch.org/docs/stable/amp.html}}.
\newblock
\newblock
\shownote{Accessed: 2025‑07‑09}.


\bibitem[Qi et~al\mbox{.}(2023)]%
        {qi2023zero}
\bibfield{author}{\bibinfo{person}{Penghui Qi}, \bibinfo{person}{Xinyi Wan}, \bibinfo{person}{Guangxing Huang}, {and} \bibinfo{person}{Min Lin}.} \bibinfo{year}{2023}\natexlab{}.
\newblock \showarticletitle{Zero bubble pipeline parallelism}.
\newblock \bibinfo{journal}{\emph{arXiv preprint arXiv:2401.10241}} (\bibinfo{year}{2023}).
\newblock


\bibitem[Rasley et~al\mbox{.}(2020)]%
        {rasley2020deepspeed}
\bibfield{author}{\bibinfo{person}{Jeff Rasley}, \bibinfo{person}{Samyam Rajbhandari}, \bibinfo{person}{Olatunji Ruwase}, {and} \bibinfo{person}{Yuxiong He}.} \bibinfo{year}{2020}\natexlab{}.
\newblock \showarticletitle{Deepspeed: System optimizations enable training deep learning models with over 100 billion parameters}. In \bibinfo{booktitle}{\emph{Proceedings of the 26th ACM SIGKDD international conference on knowledge discovery \& data mining}}. \bibinfo{pages}{3505--3506}.
\newblock


\bibitem[Sanders et~al\mbox{.}(2009)]%
        {sanders2009two}
\bibfield{author}{\bibinfo{person}{Peter Sanders}, \bibinfo{person}{Jochen Speck}, {and} \bibinfo{person}{Jesper~Larsson Tr{\"a}ff}.} \bibinfo{year}{2009}\natexlab{}.
\newblock \showarticletitle{Two-tree algorithms for full bandwidth broadcast, reduction and scan}.
\newblock \bibinfo{journal}{\emph{Parallel Comput.}} \bibinfo{volume}{35}, \bibinfo{number}{12} (\bibinfo{year}{2009}), \bibinfo{pages}{581--594}.
\newblock


\bibitem[Sato(2023)]%
        {sato2023optical}
\bibfield{author}{\bibinfo{person}{Ken-ichi Sato}.} \bibinfo{year}{2023}\natexlab{}.
\newblock \showarticletitle{Optical switching will innovate intra data center networks [Invited Tutorial]}.
\newblock \bibinfo{journal}{\emph{Journal of Optical Communications and Networking}} \bibinfo{volume}{16}, \bibinfo{number}{1} (\bibinfo{year}{2023}), \bibinfo{pages}{A1--A23}.
\newblock


\bibitem[Shah et~al\mbox{.}(2023)]%
        {taccl}
\bibfield{author}{\bibinfo{person}{Aashaka Shah}, \bibinfo{person}{Vijay Chidambaram}, \bibinfo{person}{Meghan Cowan}, \bibinfo{person}{Saeed Maleki}, \bibinfo{person}{Madan Musuvathi}, \bibinfo{person}{Todd Mytkowicz}, \bibinfo{person}{Jacob Nelson}, \bibinfo{person}{Olli Saarikivi}, {and} \bibinfo{person}{Rachee Singh}.} \bibinfo{year}{2023}\natexlab{}.
\newblock \showarticletitle{{TACCL}: Guiding Collective Algorithm Synthesis using Communication Sketches}. In \bibinfo{booktitle}{\emph{20th USENIX Symposium on Networked Systems Design and Implementation (NSDI 23)}}. \bibinfo{publisher}{USENIX Association}, \bibinfo{address}{Boston, MA}, \bibinfo{pages}{593--612}.
\newblock
\showISBNx{978-1-939133-33-5}
\urldef\tempurl%
\url{https://www.usenix.org/conference/nsdi23/presentation/shah}
\showURL{%
\tempurl}


\bibitem[Shoeybi et~al\mbox{.}(2019)]%
        {shoeybi2019megatron}
\bibfield{author}{\bibinfo{person}{Mohammad Shoeybi}, \bibinfo{person}{Mostofa Patwary}, \bibinfo{person}{Raul Puri}, \bibinfo{person}{Patrick LeGresley}, \bibinfo{person}{Jared Casper}, {and} \bibinfo{person}{Bryan Catanzaro}.} \bibinfo{year}{2019}\natexlab{}.
\newblock \showarticletitle{Megatron-lm: Training multi-billion parameter language models using model parallelism}.
\newblock \bibinfo{journal}{\emph{arXiv preprint arXiv:1909.08053}} (\bibinfo{year}{2019}).
\newblock


\bibitem[Shrivastav et~al\mbox{.}(2019)]%
        {shoal}
\bibfield{author}{\bibinfo{person}{Vishal Shrivastav}, \bibinfo{person}{Asaf Valadarsky}, \bibinfo{person}{Hitesh Ballani}, \bibinfo{person}{Paolo Costa}, \bibinfo{person}{Ki~Suh Lee}, \bibinfo{person}{Han Wang}, \bibinfo{person}{Rachit Agarwal}, {and} \bibinfo{person}{Hakim Weatherspoon}.} \bibinfo{year}{2019}\natexlab{}.
\newblock \showarticletitle{Shoal: A Network Architecture for Disaggregated Racks}. In \bibinfo{booktitle}{\emph{16th USENIX Symposium on Networked Systems Design and Implementation (NSDI 19)}}. \bibinfo{publisher}{USENIX Association}, \bibinfo{address}{Boston, MA}, \bibinfo{pages}{255--270}.
\newblock
\showISBNx{978-1-931971-49-2}
\urldef\tempurl%
\url{https://www.usenix.org/conference/nsdi19/presentation/shrivastav}
\showURL{%
\tempurl}


\bibitem[Singh et~al\mbox{.}(2015)]%
        {jupiter-rising}
\bibfield{author}{\bibinfo{person}{Arjun Singh}, \bibinfo{person}{Joon Ong}, \bibinfo{person}{Amit Agarwal}, \bibinfo{person}{Glen Anderson}, \bibinfo{person}{Ashby Armistead}, \bibinfo{person}{Roy Bannon}, \bibinfo{person}{Seb Boving}, \bibinfo{person}{Gaurav Desai}, \bibinfo{person}{Bob Felderman}, \bibinfo{person}{Paulie Germano}, \bibinfo{person}{Anand Kanagala}, \bibinfo{person}{Jeff Provost}, \bibinfo{person}{Jason Simmons}, \bibinfo{person}{Eiichi Tanda}, \bibinfo{person}{Jim Wanderer}, \bibinfo{person}{Urs H\"{o}lzle}, \bibinfo{person}{Stephen Stuart}, {and} \bibinfo{person}{Amin Vahdat}.} \bibinfo{year}{2015}\natexlab{}.
\newblock \showarticletitle{Jupiter Rising: A Decade of Clos Topologies and Centralized Control in Google's Datacenter Network}.
\newblock \bibinfo{journal}{\emph{SIGCOMM Comput. Commun. Rev.}} \bibinfo{volume}{45}, \bibinfo{number}{4} (\bibinfo{date}{aug} \bibinfo{year}{2015}), \bibinfo{pages}{183–197}.
\newblock
\showISSN{0146-4833}
\urldef\tempurl%
\url{https://doi.org/10.1145/2829988.2787508}
\showDOI{\tempurl}


\bibitem[Singla et~al\mbox{.}(2014)]%
        {singla2014high}
\bibfield{author}{\bibinfo{person}{Ankit Singla}, \bibinfo{person}{P~Brighten Godfrey}, {and} \bibinfo{person}{Alexandra Kolla}.} \bibinfo{year}{2014}\natexlab{}.
\newblock \showarticletitle{High throughput data center topology design}. In \bibinfo{booktitle}{\emph{11th USENIX Symposium on Networked Systems Design and Implementation (NSDI 14)}}. \bibinfo{pages}{29--41}.
\newblock


\bibitem[Singla et~al\mbox{.}(2012)]%
        {jellyfish}
\bibfield{author}{\bibinfo{person}{Ankit Singla}, \bibinfo{person}{Chi-Yao Hong}, \bibinfo{person}{Lucian Popa}, {and} \bibinfo{person}{P.~Brighten Godfrey}.} \bibinfo{year}{2012}\natexlab{}.
\newblock \showarticletitle{Jellyfish: Networking Data Centers Randomly}. In \bibinfo{booktitle}{\emph{9th USENIX Symposium on Networked Systems Design and Implementation (NSDI 12)}}. \bibinfo{publisher}{USENIX Association}, \bibinfo{address}{San Jose, CA}, \bibinfo{pages}{225--238}.
\newblock
\showISBNx{978-931971-92-8}
\urldef\tempurl%
\url{https://www.usenix.org/conference/nsdi12/technical-sessions/presentation/singla}
\showURL{%
\tempurl}


\bibitem[Sreenilayam et~al\mbox{.}(2019)]%
        {sreenilayam2019fast}
\bibfield{author}{\bibinfo{person}{Sithara~P Sreenilayam}, \bibinfo{person}{Dermot Brabazon}, {and} \bibinfo{person}{Yuri~P Panarin}.} \bibinfo{year}{2019}\natexlab{}.
\newblock \showarticletitle{Fast ferroelectric liquid crystal based optical switch: simulation and experiments}.
\newblock \bibinfo{journal}{\emph{Crystals}} \bibinfo{volume}{9}, \bibinfo{number}{8} (\bibinfo{year}{2019}), \bibinfo{pages}{388}.
\newblock


\bibitem[Tazi et~al\mbox{.}(2025)]%
        {ultrascale_playbook}
\bibfield{author}{\bibinfo{person}{Nouamane Tazi}, \bibinfo{person}{Ferdinand Mom}, \bibinfo{person}{Haojun Zhao}, \bibinfo{person}{Phuc Nguyen}, \bibinfo{person}{Mohamed Mekkouri}, \bibinfo{person}{Leandro Werra}, {and} \bibinfo{person}{Thomas Wolf}.} \bibinfo{year}{2025}\natexlab{}.
\newblock \bibinfo{title}{The Ultra-Scale Playbook: Training LLMs on GPU Clusters}.
\newblock \bibinfo{howpublished}{\url{https://huggingface.co/spaces/nanotron/ultrascale-playbook}}.
\newblock
\newblock
\shownote{Accessed: 2025-05-16}.


\bibitem[{Telescent Inc.}(nd)]%
        {telescent_products}
\bibfield{author}{\bibinfo{person}{{Telescent Inc.}}} \bibinfo{year}{n.d.}\natexlab{}.
\newblock \bibinfo{title}{Products | Telescent}.
\newblock \bibinfo{howpublished}{\url{https://www.telescent.com/products}}.
\newblock
\newblock
\shownote{Accessed: 2025-07-01}.


\bibitem[Thakur and Gropp(2003)]%
        {thakur2003improving}
\bibfield{author}{\bibinfo{person}{Rajeev Thakur} {and} \bibinfo{person}{William~D Gropp}.} \bibinfo{year}{2003}\natexlab{}.
\newblock \showarticletitle{Improving the performance of collective operations in MPICH}. In \bibinfo{booktitle}{\emph{European Parallel Virtual Machine/Message Passing Interface Users’ Group Meeting}}. Springer, \bibinfo{pages}{257--267}.
\newblock


\bibitem[Wang et~al\mbox{.}(2010)]%
        {cthrough}
\bibfield{author}{\bibinfo{person}{Guohui Wang}, \bibinfo{person}{David~G. Andersen}, \bibinfo{person}{Michael Kaminsky}, \bibinfo{person}{Konstantina Papagiannaki}, \bibinfo{person}{T.S.~Eugene Ng}, \bibinfo{person}{Michael Kozuch}, {and} \bibinfo{person}{Michael Ryan}.} \bibinfo{year}{2010}\natexlab{}.
\newblock \showarticletitle{C-Through: Part-Time Optics in Data Centers}. In \bibinfo{booktitle}{\emph{Proceedings of the ACM SIGCOMM 2010 Conference}} (New Delhi, India) \emph{(\bibinfo{series}{SIGCOMM '10})}. \bibinfo{publisher}{Association for Computing Machinery}, \bibinfo{address}{New York, NY, USA}, \bibinfo{pages}{327–338}.
\newblock
\showISBNx{9781450302012}
\urldef\tempurl%
\url{https://doi.org/10.1145/1851182.1851222}
\showDOI{\tempurl}


\bibitem[Wang et~al\mbox{.}(2024)]%
        {wang2024rail}
\bibfield{author}{\bibinfo{person}{Weiyang Wang}, \bibinfo{person}{Manya Ghobadi}, \bibinfo{person}{Kayvon Shakeri}, \bibinfo{person}{Ying Zhang}, {and} \bibinfo{person}{Naader Hasani}.} \bibinfo{year}{2024}\natexlab{}.
\newblock \showarticletitle{Rail-only: A low-cost high-performance network for training LLMs with trillion parameters}. In \bibinfo{booktitle}{\emph{2024 IEEE Symposium on High-Performance Interconnects (HOTI)}}. IEEE, \bibinfo{pages}{1--10}.
\newblock


\bibitem[Wang et~al\mbox{.}(2023)]%
        {wang2023topoopt}
\bibfield{author}{\bibinfo{person}{Weiyang Wang}, \bibinfo{person}{Moein Khazraee}, \bibinfo{person}{Zhizhen Zhong}, \bibinfo{person}{Manya Ghobadi}, \bibinfo{person}{Zhihao Jia}, \bibinfo{person}{Dheevatsa Mudigere}, \bibinfo{person}{Ying Zhang}, {and} \bibinfo{person}{Anthony Kewitsch}.} \bibinfo{year}{2023}\natexlab{}.
\newblock \showarticletitle{$\{$TopoOpt$\}$: Co-optimizing network topology and parallelization strategy for distributed training jobs}. In \bibinfo{booktitle}{\emph{20th USENIX Symposium on Networked Systems Design and Implementation (NSDI 23)}}. \bibinfo{pages}{739--767}.
\newblock


\bibitem[Wu et~al\mbox{.}(2023)]%
        {sipac}
\bibfield{author}{\bibinfo{person}{Zhenguo Wu}, \bibinfo{person}{Liang~Yuan Dai}, \bibinfo{person}{Ziyi Zhu}, \bibinfo{person}{Asher Novick}, \bibinfo{person}{Madeleine Glick}, {and} \bibinfo{person}{Keren Bergman}.} \bibinfo{year}{2023}\natexlab{}.
\newblock \showarticletitle{SiP Architecture For Accelerating Collective Communication in Distributed Deep Learning}. In \bibinfo{booktitle}{\emph{2023 Optical Fiber Communications Conference and Exhibition (OFC)}}. \bibinfo{pages}{1--3}.
\newblock
\urldef\tempurl%
\url{https://doi.org/10.1364/OFC.2023.W1G.1}
\showDOI{\tempurl}


\bibitem[Xing et~al\mbox{.}(2015)]%
        {xing2015petuum}
\bibfield{author}{\bibinfo{person}{Eric~P Xing}, \bibinfo{person}{Qirong Ho}, \bibinfo{person}{Wei Dai}, \bibinfo{person}{Jin-Kyu Kim}, \bibinfo{person}{Jinliang Wei}, \bibinfo{person}{Seunghak Lee}, \bibinfo{person}{Xun Zheng}, \bibinfo{person}{Pengtao Xie}, \bibinfo{person}{Abhimanu Kumar}, {and} \bibinfo{person}{Yaoliang Yu}.} \bibinfo{year}{2015}\natexlab{}.
\newblock \showarticletitle{Petuum: A new platform for distributed machine learning on big data}. In \bibinfo{booktitle}{\emph{Proceedings of the 21th ACM SIGKDD International Conference on Knowledge Discovery and Data Mining}}. \bibinfo{pages}{1335--1344}.
\newblock


\bibitem[Yeluri(2023)]%
        {Yeluri2023_power_consumption}
\bibfield{author}{\bibinfo{person}{Sharada Yeluri}.} \bibinfo{year}{2023}\natexlab{}.
\newblock \bibinfo{booktitle}{\emph{Optimizing Power Consumption in High‑End Routers}}.
\newblock


\bibitem[Zhao et~al\mbox{.}(2023)]%
        {zhao2023pytorch}
\bibfield{author}{\bibinfo{person}{Yanli Zhao}, \bibinfo{person}{Andrew Gu}, \bibinfo{person}{Rohan Varma}, \bibinfo{person}{Liang Luo}, \bibinfo{person}{Chien-Chin Huang}, \bibinfo{person}{Min Xu}, \bibinfo{person}{Less Wright}, \bibinfo{person}{Hamid Shojanazeri}, \bibinfo{person}{Myle Ott}, \bibinfo{person}{Sam Shleifer}, {et~al\mbox{.}}} \bibinfo{year}{2023}\natexlab{}.
\newblock \showarticletitle{Pytorch fsdp: experiences on scaling fully sharded data parallel}.
\newblock \bibinfo{journal}{\emph{arXiv preprint arXiv:2304.11277}} (\bibinfo{year}{2023}).
\newblock


\bibitem[Zu et~al\mbox{.}(2024)]%
        {tpuresilience}
\bibfield{author}{\bibinfo{person}{Yazhou Zu}, \bibinfo{person}{Alireza Ghaffarkhah}, \bibinfo{person}{Hoang-Vu Dang}, \bibinfo{person}{Brian Towles}, \bibinfo{person}{Steven Hand}, \bibinfo{person}{Safeen Huda}, \bibinfo{person}{Adekunle Bello}, \bibinfo{person}{Alexander Kolbasov}, \bibinfo{person}{Arash Rezaei}, \bibinfo{person}{Dayou Du}, \bibinfo{person}{Steve Lacy}, \bibinfo{person}{Hang Wang}, \bibinfo{person}{Aaron Wisner}, \bibinfo{person}{Chris Lewis}, {and} \bibinfo{person}{Henri Bahini}.} \bibinfo{year}{2024}\natexlab{}.
\newblock \showarticletitle{Resiliency at Scale: Managing {Google{\textquoteright}s} {TPUv4} Machine Learning Supercomputer}. In \bibinfo{booktitle}{\emph{21st USENIX Symposium on Networked Systems Design and Implementation (NSDI 24)}}. \bibinfo{pages}{761--774}.
\newblock


\end{thebibliography}

\end{document}